\documentclass[aps,prb,twocolumn,showpacs]{revtex4}

\usepackage{latexsym}
\usepackage{graphicx}
\usepackage{times,psfrag,subfigure}
\usepackage{amsmath}
\usepackage{dcolumn}
\usepackage{latexsym,amsmath,amssymb,bm,euscript}
\usepackage{afterpage}
\newcommand{\be}{\begin{equation}}
\newcommand{\ee}{\end{equation}}
\newcommand{\bea}{\begin{eqnarray}}
\newcommand{\eea}{\end{eqnarray}}
\newcommand{\divo}{\operatorname{div}}
\newcommand{\curl}{\operatorname{curl}}

\begin{document}

\title{Coulomb gas transitions in three-dimensional classical dimer models}
\date{\today}

\author{Gang Chen}
\affiliation{Physics Department, 
University of California, Santa Barbara, CA 93106}
\author{Jan Gukelberger}
\affiliation{Physics Department, 
University of California, Santa Barbara, CA 93106}
\affiliation{Microsoft Research, Station Q, 
University of California, Santa Barbara, CA 93106}
\author{Simon Trebst}
\affiliation{Microsoft Research, Station Q, 
University of California, Santa Barbara, CA 93106}
\author{Fabien Alet}
\affiliation{Laboratoire de Physique Th\'eorique, Universit\'e de Toulouse,
 UPS, (IRSAMC), 31062 Toulouse, France}
\affiliation{CNRS, LPT (IRSAMC), 31062 Toulouse, France}
\author{Leon Balents}
\affiliation{Kavli Institute for Theoretical Physics, 
University of California, Santa Barbara, CA 93106}

\begin{abstract}
Close-packed, classical dimer models on three-dimensional, bipartite lattices harbor a 
Coulomb phase with power-law correlations at infinite temperature. 
Here, we discuss the nature of the thermal phase transition out of this Coulomb phase 
for a variety of dimer models which energetically favor crystalline dimer states with columnar ordering. 
For a family of these models we find a direct thermal transition from the Coulomb phase to the
dimer crystal. While some systems exhibit (strong) first-order transitions in correspondence
with the Landau-Ginzburg-Wilson paradigm, we also find clear numerical evidence for continuous 
transitions. 
A second family of models undergoes two consecutive thermal transitions with an intermediate
paramagnetic phase separating the Coulomb phase from the dimer crystal.
We can describe all of these phase transitions in one unifying framework of candidate field theories 
with two complex Ginzburg-Landau fields coupled to a U(1) gauge field.
We derive the symmetry-mandated Ginzburg-Landau actions in these field variables 
for the various dimer models and discuss implications for their respective phase transitions. 
\end{abstract}


\maketitle

\section{Introduction}
\label{sec:intro}

{\sl Constraints} are a pervasive feature of strongly correlated
systems.  For instance, in Mott insulators, a large atomic Coulomb
repulsion effectively constrains the charge of each ion to be fixed,
while still allowing spin and orbital fluctuations.  This situation, in
which the dominant terms in the Hamiltonian impose constraints, but
local fluctuations remain strong, provides a challenge to physical
understanding.  In frustrated magnets, it is common to observe a
``cooperative paramagnetic'' regime, in which the dominant exchange
interactions impose strong constraints on the spin configurations, but
the spins still manage to remain strongly fluctuating.  The ``spin ice''
materials, in which rare earth Ising moments locally satisfy Pauling's
ice rules, provide a prominent and beautiful set of examples, which have
stimulated a rich interplay between theory and experiment.  Constraints
on the spin phase space have been implicated in the physics of diverse
other magnetic materials, such as the spinel chromites\cite{chromites}
and the A-site diamond antiferromagnetic spinels\cite{bergman:nature}.

Generally, residual interactions, subdominant to those responsible for
the constraints, lead to a quenching of the remaining fluctuations.  To
quantify this, we may associate the dominant interactions with a
temperature, $T_0$, below which the constraints are well satisfied and
the system is highly correlated.  We will assume that fluctuations
amongst the constrained states are removed at another temperature
$T_c\ll T_0$, which is determined by subdominant effects.  Often this
quenching of the constrained fluctuations is associated with a symmetry
breaking, such as magnetic ordering or lattice deformation.  This phase
transition occurs in a very different environment from conventional
order-disorder transitions, in which the high temperature phase is a
weakly correlated paramagnet.  Here, the strong constraints imply strong
correlations in the cooperative paramagnet.  It has recently been
appreciated that such correlations can drastically affect phase
transition(s).\cite{bergman:prb2006,alet:prl2006,Pickles,chamon} Transitions can be
induced where none would otherwise be present, and furthermore,
symmetry-mandated transitions may be modified from their usual
Landau-Ginzburg-Wilson (LGW) universality classes.  

In this paper, we explore these phenomena in a large set of classical
dimer models on the cubic lattice.  Such dimer models are defined by a
constrained phase space consisting of close-packed dimer coverings, in
which dimers occupy (some) links of the lattice, and the constraint is
that each site is overlapped by one and only one dimer.  In these
models, the constraint is exactly satisfied, corresponding to the limit
$T_0 \rightarrow \infty$.  We expect that more realistic models are well
approximated by this situation provided $T_c \ll T_0$.  For models with
a gap (of order $O(k_B T_0)$) to states violating the constraint (as in spin
ice), the approximation is in fact exponentially good, since violations
of the constraint occur with an Arrhenius probability 
$\propto \exp(-T_0/T_c)$.  

In the cubic dimer models we study, a great deal is understood about the
nature of the constraint induced correlations \cite{huse:prl2003}, which have a power-law
form.  This can be cast (see Sec.~\ref{sec:field}) into a sort of
pseudo-dipolar form, leading to the name ``Coulomb phase'' for the high
temperature $T>T_c$ region.  Moreover, the dipolar correlations can be
identified with an emergent Coulomb gauge field, similar to that
appearing in true electromagnetism.  Coulomb phases arise in a variety
of other contexts \cite{Hermele:prb04,PyrochloreCoulombPhases}; 
for instance, the ice rules constraint related to spin ice also leads to a Coulomb phase 
\cite{PyrochloreCoulombPhases,EarlyWorkCoulombPhases1,EarlyWorkCoulombPhases2,EarlyWorkCoulombPhases3}.  
For these types of systems, the
gauge description has lead to some theoretical progress in understanding
the consequent unconventional criticality.   In such cases, the
transitions are expected to involve dual ``monopole'' fields,
which couple to the emergent gauge field and carry the associated gauge
charge.\cite{bergman:prb2006,Pickles}\  The full field theory therefore has a
multicomponent Ginzburg-Landau form.  The monopoles are ``fractional'' degrees
of freedom in the sense that the symmetry-breaking order parameters (if
any) are composites of these fields.  We outline the derivation of this
result in Sec.~\ref{sec:field}.  

Though this construction of non-LGW critical theories has been discussed
in some isolated instances previously, an unequivocal verification of
the theory has proved difficult.  In particular, the numerical experiments
in Refs.~\onlinecite{alet:prl2006,misguich:prb08}, while providing evidence 
for unconventional criticality, are not in good quantitative agreement with the 
theoretical expectations.
In particular, the numerical estimates  \cite{alet:prl2006} of various critical exponents, 
e.g. $\nu =0.50(4)$, $\alpha = 0.56(7)$ and $\eta = -0.02(5)$,
are indicative of an unexplained tricritical behavior.

In this paper, we perform a much more systematic
investigation of a range of dimer models, in order to test the
theoretical picture on a grander scale, in which many  {\sl qualitative}
comparisons are possible.  We find that the gauge theory does an
excellent job on these qualitative tests, providing understanding of the
numerical results for nearly all cases.  


\subsection{Outline of models and results}
\label{sec:model}

\begin{figure}[t]
\includegraphics[width=\columnwidth]{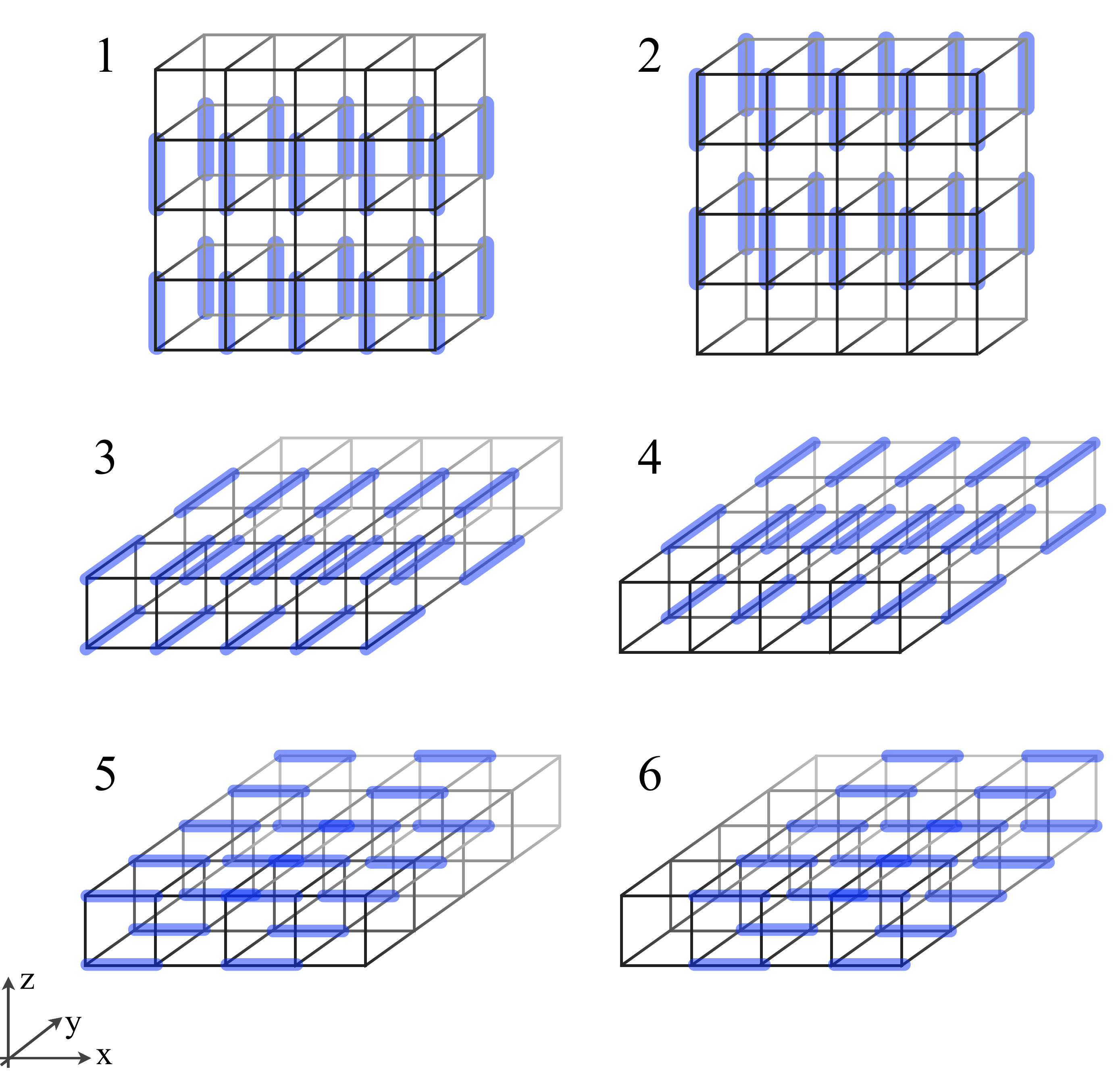}
\caption{
  (color online) The six dimer coverings with columnar dimer ordering that maximize
  the number of parallel dimers on neighboring bonds of the cubic lattice. 
  Our family of classical dimer models energetically favors these ordering patterns
  with the `6-GS' model favoring all six states, the `4-GS' model favoring states $3-6$, 
  the `2-GS' model favoring states $1-2$ and the `1-GS' model favoring a single state only.
  Further variations are discussed in the text.
}
\label{Fig:DimerModels}
\end{figure}

Before going into a detailed discussion we first provide a brief overview of the models we will
study and our main results. We start by introducing a family of close-packed, classical dimer 
models on the cubic lattice.  The elementary degrees of freedom in these models are hard-core 
dimers occupying the bonds of the cubic lattice with the constraint that every site in the lattice 
is part of exactly one such dimer. We further introduce a (potential) energy scale favoring dimer coverings with parallel dimers on neighboring bonds.
At low temperature these models exhibit  long range columnar order of the dimers. In general, 
there are six distinct columnar ordering patterns  that can maximize the number of parallel 
dimers on neighboring bonds as illustrated in  Fig.~\ref{Fig:DimerModels}.

The main distinction between the various models we will consider then comes 
from a selection of a subset of these energetically favored ordering patterns.
The Hamiltonian of a parent model that favors all six possible ordering patterns 
and which we therefore call the `6-GS' model is given by
\be
{\mathcal H}_{\text{6-GS}} = - \sum_{\substack{\square}} (n_{=} + n_{//} + n_{||}) 
\;.
\label{eq:dimerH}
\ee
Here $n_{=}$, $n_{//}$, and $n_{||}$ count the number of parallel dimers 
on neighboring bonds along the $x,y,z$ lattice directions and the sum runs
over all square plaquettes of the cubic lattice.

Our first family of dimer models selects particular subsets of four, two and one 
columnar dimer coverings as ground states: 
The `4-GS' model favors the four columnar states in $x$ and $y$ directions, 
e.g. states $3-6$ in Fig.~\ref{Fig:DimerModels}, and is described by the Hamiltonian
\be
{\mathcal H}_{\text{4-GS}} = - \sum_{\substack{\square}} (n_{=} + n_{//}) \;.
\label{eq:4gs}
\ee
The `2-GS' model favors columnar orderings only along one lattice direction, 
say the $z$ direction, e.g. states $1-2$ in Fig.~\ref{Fig:DimerModels}, with Hamiltonian
\be
{\mathcal H}_{\text{2-GS}} = - \sum_{\substack{\square}} n_{||} \;.
\label{eq:2gs}
\ee
While these two models have somewhat different ground-state properties, they will turn 
out to exhibit rather similar physics.  More specifically they share the same set of symmetries, 
which we will discuss in detail in section \ref{sec:duality}.
The last member of this first family of models is  the `1-GS' model which singles out one of 
the six columnar orderings as sole ground state.  We choose one of the columnar orderings 
in the $z$ direction and define $n^{e/o}_{||}$ to be  the number of plaquettes with 
parallel dimers on neighboring even or odd bonds, respectively. 
Then the Hamiltonian of the 1-GS model  can be written as
\be
{\mathcal H}_{\text{1-GS}} = - \sum_{\substack{\square}} n^e_{||} \;. 
\label{eq:1gs}
\ee

A common feature of all four models introduced above is that they all undergo a {\em direct}
thermal transition between the Coulomb gas phase at high temperature and a conventional
long-range ordered state at low temperature as we will discuss in Sec.~\ref{sec:numerics}. 
Our parent Hamiltonian, the 6-GS model, has first been studied in Ref.~\onlinecite{alet:prl2006}, 
where based on an extensive numerical analysis the authors argue that this model undergoes a
{\sl continuous} thermal transition with the system spontaneously selecting one of the six columnar 
ordering patterns at the transition out of the Coulomb phase.
Our present numerical analysis for the other three models indicates that the order of their respective transitions is (strongly) first order for the 2-GS and 4-GS models, while we find strong evidence that the transition of the 1-GS model is again continuous.

In Sec.~\ref{sec:interpolation} we provide further numerical evidence for the continuous nature of the 
1-GS model by embedding this transition into a line of continuous transitions for systems of identical symmetry. 
The latter is achieved by studying a continuous interpolation of the 1-GS and 2-GS model 
in Sec.~\ref{sec:interpolation-1GS}. A similar approach interpolating the 4-GS to 6-GS model is given in Sec.~\ref{sec:interpolation-6GS}.

To understand the different phase transitions found numerically, we develop candidate field theories 
in Sec.~\ref{sec:field}. To this end, we first rewrite the dimer models in terms of a compact U(1) gauge theory. A subsequent duality transformation then allows to make the monopole excitations in this
gauge theory explicit and describe the phase transitions as Higgs confinement transitions driven by the condensation of monopoles. Finally, the candidate field theories are derived in terms of two complex fields (a $CP^1$ field) coupled to a U(1) gauge field. The symmetry-mandated Landau-Ginzburg actions for the various dimer models are given in Sec.~\ref{sec:action} and implications for the phase transitions
discussed. 
In particular, this analysis suggests that the 1-GS model undergoes a  continuous transition in the 3D
inverted XY universality class, while the 2-GS model should undergo a first-order transition. While analytically inferring the nature of the transition for  the 4-GS and 6-GS model is somewhat more 
delicate as discussed in Sec.~\ref{sec:action}, we find an overall good agreement with the numerical results.

In Sec.~\ref{sec:variation} we discuss a second family of dimer models with the common characteristic 
that there are two subsequent thermal transitions between the Coulomb phase and the dimer crystal. 
As a representative model we discuss the so-called `xy' model in some detail, which favors one
columnar ordering pattern along the $x$ and $y$ lattice directions, respectively.  
We find that the system first undergoes a continuous Higgs transition out of the Coulomb phase, 
which again is in the 3D inverted XY universality class, and subsequently a first-order ``spin-flop" 
transition to the dimer crystal. 

We give some concluding remarks in Sec.~\ref{sec:discuss} on the general interest of these classical dimer models in the search of non-LGW phase transitions.


\section{Monte Carlo Simulations}
\label{sec:numerics}

\subsection{Overview}

We first summarize some characteristic numerical results for the thermal transitions obtained 
from extensive Monte Carlo simulations.
The classical nature of our models not only commands to use an efficient stochastic algorithm
to traverse the space of dimer coverings, but also allows for non-local update schemes such as 
the worm  algorithm \cite{sandvik:prb2006,alet:pre2006}.
The latter performs an update by flipping a whole sequence of dimers 
when moving from one dimer covering to another one, which drastically reduces the problem
of critical slowing down close to phase transitions. 
We used this algorithm on samples with $N=L^3$ sites up to $256^3$ -- 
sizes which sometimes turn out to be necessary to ascertain the nature of a phase transition. 

The update scheme of the worm algorithm further allows to sample the behavior of two test 
monomers embedded in the dimer coverings. In particular, the update is performed by initially 
breaking up an arbitrary dimer into a pair of monomers and then moving one monomer across 
the lattice by flipping dimers along a string or `worm' until it can be recombined with the other
monomer into a newly formed dimer. 
This construction can be used to reveal the confining properties of the low-temperature phases 
in our dimer models. To this end, we define the monomer `confinement length' $\xi^2(T)$
as the (squared) average distance between the two test monomers, which we rescale by the 
expectation value $(L^2+2)/4$ for deconfined monomers moving freely on the lattice for a 
finite cube of even linear extent $L$ (and periodic boundary conditions).

\begin{figure}[htp]
\includegraphics[width=\columnwidth]{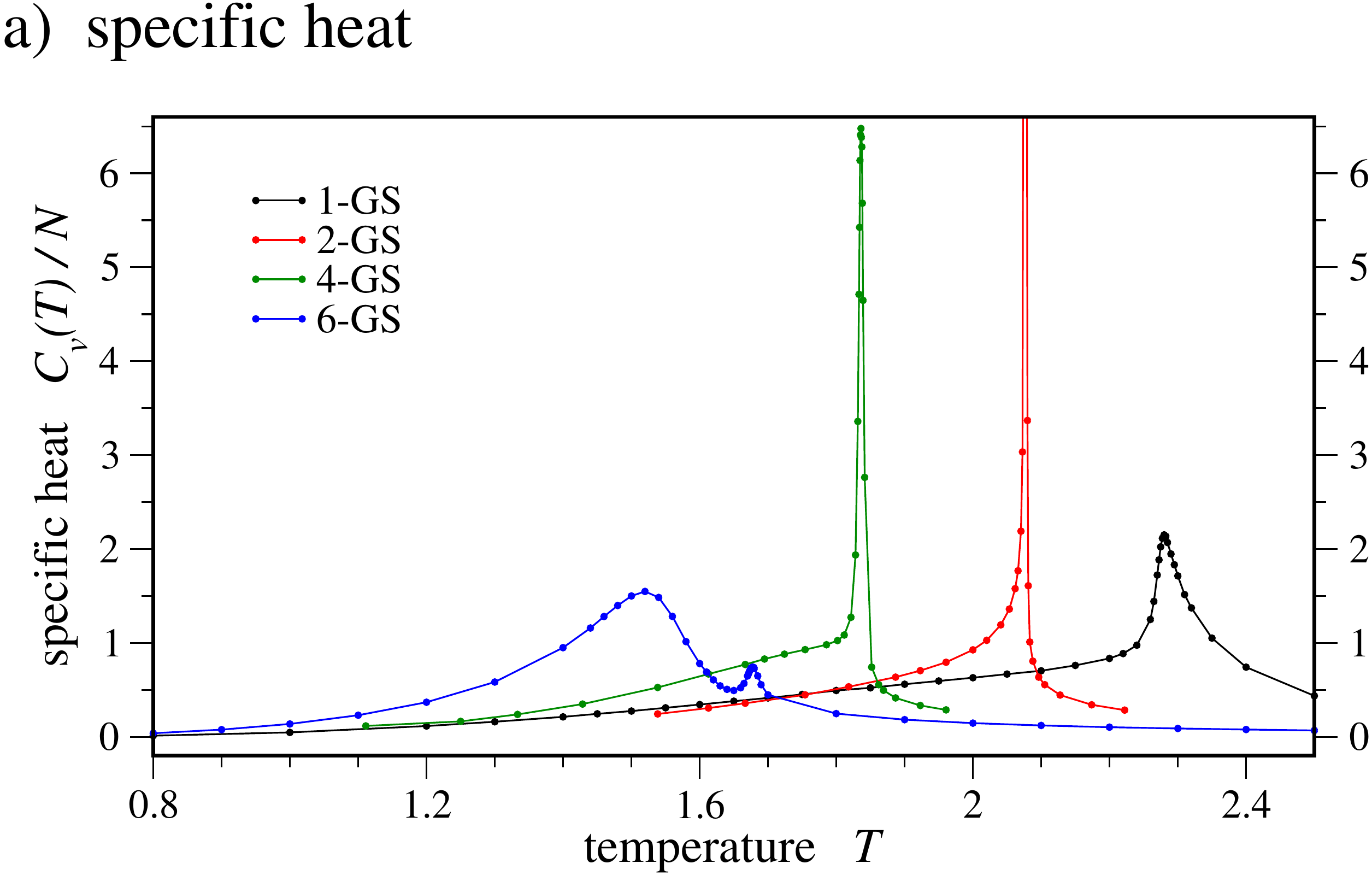}
\vskip 4mm
\includegraphics[width=\columnwidth]{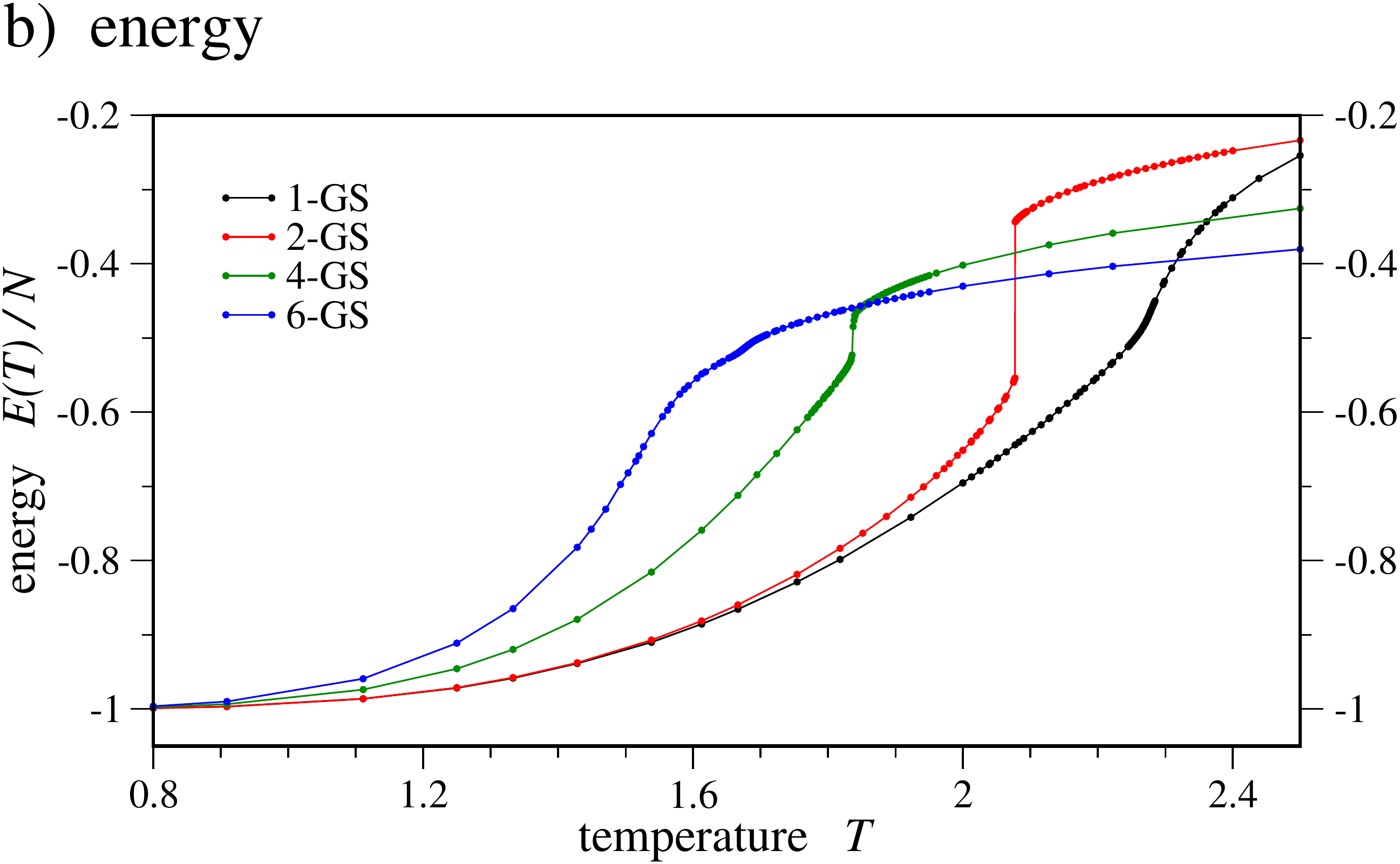}
\vskip 4mm
\includegraphics[width=\columnwidth]{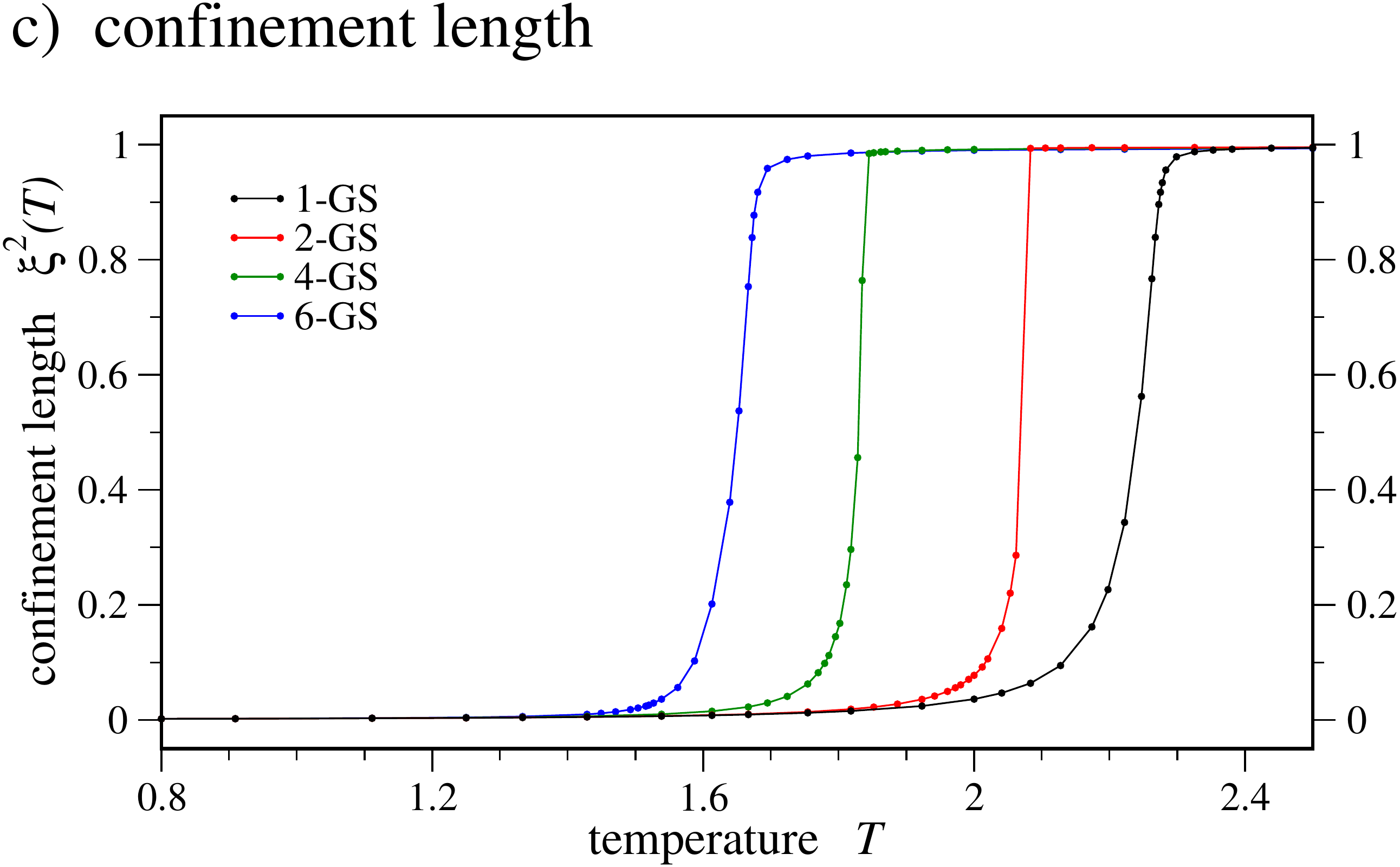}
\caption{
  (color online) 
  Overview of numerical results for the dimer models with one (1-GS), two (2-GS), four (4-GS) 
  and six (6-GS) columnar ground states: a) the specific heat per site $C_v(T)/N$, b)
  the energy per site $E(T)/N$, 
  and c) the monomer confinement length $\xi^2(T)$ defined in the text.
  All four models undergo a direct transition with clear thermodynamic signatures
  between the high-temperature Coulomb phase (with deconfined monomers)
  to the dimer crystal at low temperature (with confined monomers).
  The sharp, kink-like features in the energy $E(T)$ and monomer confinement length $\xi^2(T)$
  are indicative of a first-order transition for the 2-GS and 4-GS models.
  The 1-GS and 6-GS model appear to undergo continuous transitions with smooth
  features. 
  Data is shown for system size $L=48$ for the 1-GS and 6-GS models, 
  and $L=32$ for the 2-GS and 4-GS models.
}
\label{Fig:nGS-overview}
\end{figure}

We also measure thermodynamical quantities such as the internal energy $E(T)$, the specific heat 
$C_v(T)=(\langle E^2\rangle - \langle E\rangle^2)/T^2$ as well as the
stiffness $\rho$. The stiffness encodes fluctuations of dimer fluxes: $\rho=\sum_{\alpha=x,y,z}\langle \phi^2_\alpha
 \rangle/3L$, where the flux $\phi_\alpha$ is the algebraic number of dimers
crossing a plane perpendicular to the unit vector $e_\alpha$. Algebraic here
means that, given a lattice direction, we count $+1$ for a dimer going from
one sublattice to the other and $-1$ for the reverse situation. 
Fluxes $\phi_\alpha$ are conserved quantities (plane by plane) which
vanish on average for symmetry reasons.

In Fig.~\ref{Fig:nGS-overview} we plot the specific heat per site
$C_v(T)/N$, the energy per site $E(T)/N$, 
and the monomer confinement length $\xi^2(T)$ for the four models 
introduced in the previous section.

For all four models we find clear thermodynamic signatures for a direct transition
between the high-temperature Coulomb phase (with deconfined monomers)
to the dimer crystal at low temperatures (with confined monomers).
The sharp, kink-like features in the energy $E(T)$ and monomer confinement length $\xi^2(T)$
are indicative of a first-order transition for the 2-GS and 4-GS models. The first-order nature
of these transitions is fully revealed in bimodal energy histograms in the vicinity of the transition
temperature, which we will discuss in detail in section \ref{sec:interpolation}.
The 1-GS and 6-GS appear to undergo continuous transitions with smooth features in the 
energy $E(T)$ and monomer confinement length $\xi^2(T)$ and a (divergent) peak in the specific
heat $C_v(T)$. We have found no evidence of bimodal energy histograms in the vicinity of 
these transitions, as also discussed in section \ref{sec:interpolation}.

We will now turn to the individual dimer models and discuss our numerical results in more detail
in the following.

\subsection{The 1-GS model}

\begin{figure}[t]
\includegraphics[width=\columnwidth]{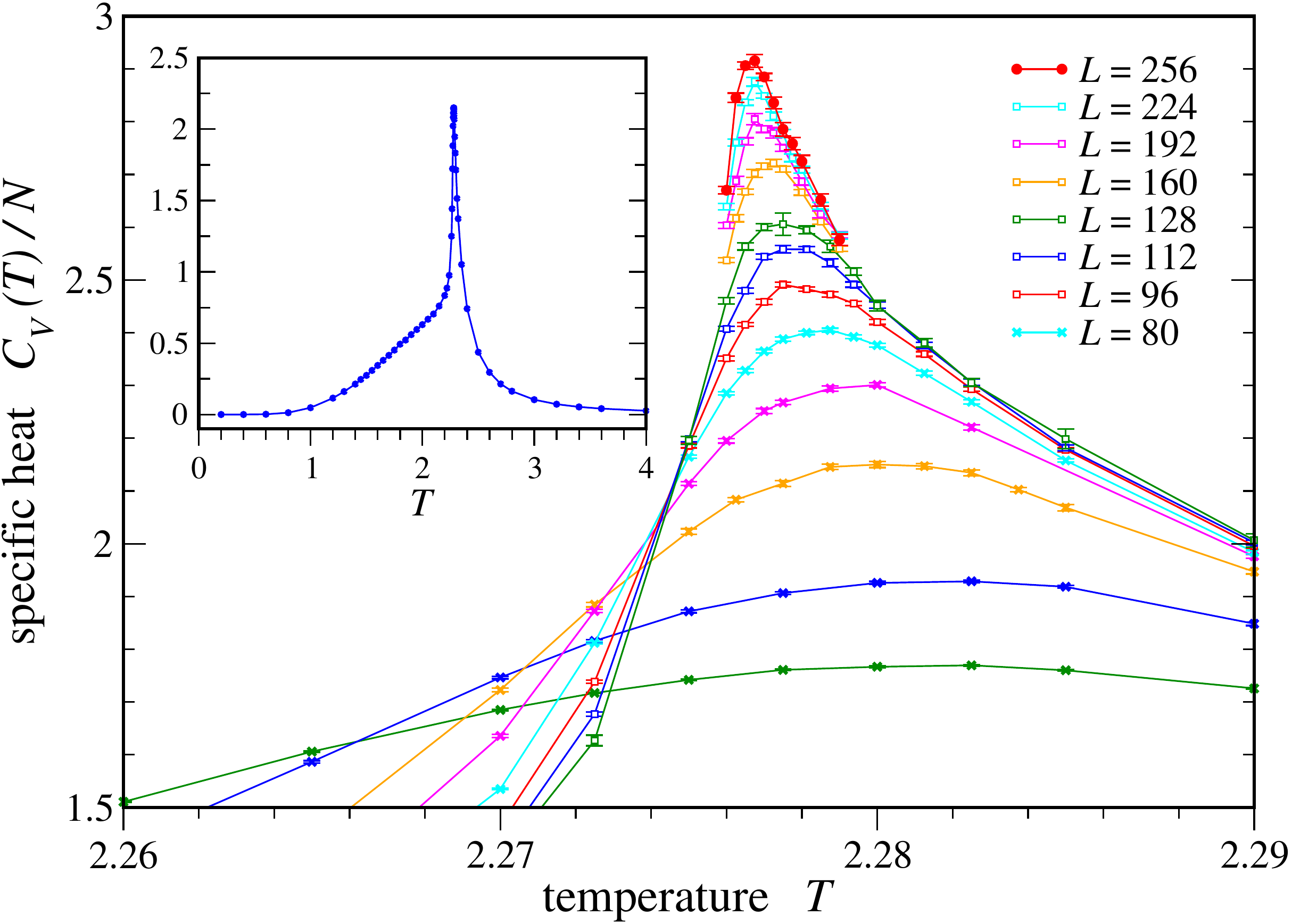}
\caption{
  (color online) 
  Specific heat per site $C_v(T)/N$ for the 1-GS model as function of temperature T
  and different system sizes $L$.
  The inset shows a specific heat scan over a wide temperature range for a sample $L=48$ system
  size. Below the peak around $T_c \approx 2.276 \pm 0.001$ there is a shoulder which
  shows no variation with system size.
}
\label{Fig:1GS-SpecHeat}
\end{figure}

We will first concentrate on the 1-GS model which energetically favors a single columnar dimer 
ordering pattern shown in Fig.~\ref{Fig:DimerModels}.
Our numerical simulations for systems with up to $256^3$ dimers clearly suggest that this model
undergoes a {\em continuous} thermal transition between the Coulomb phase and the dimer crystal.

The specific heat plotted in Fig.~\ref{Fig:1GS-SpecHeat} exhibits a peak around the transition
temperature of $T_c \approx 2.276 \pm 0.001$ that appears to diverge very slowly with $L$.
Below this peak there is a shoulder that does not show any variation with system size, 
(see inset of Fig.~\ref{Fig:1GS-SpecHeat}), and thus cannot be associated
with any long distance or critical behavior. The latter is reminiscent of
the 6-GS model \cite{alet:prl2006} which below the transition temperature
exhibits an even more pronounced shoulder (for a comparison see also Fig.~\ref{Fig:nGS-overview}).

\begin{figure}[t]
\includegraphics[width=\columnwidth]{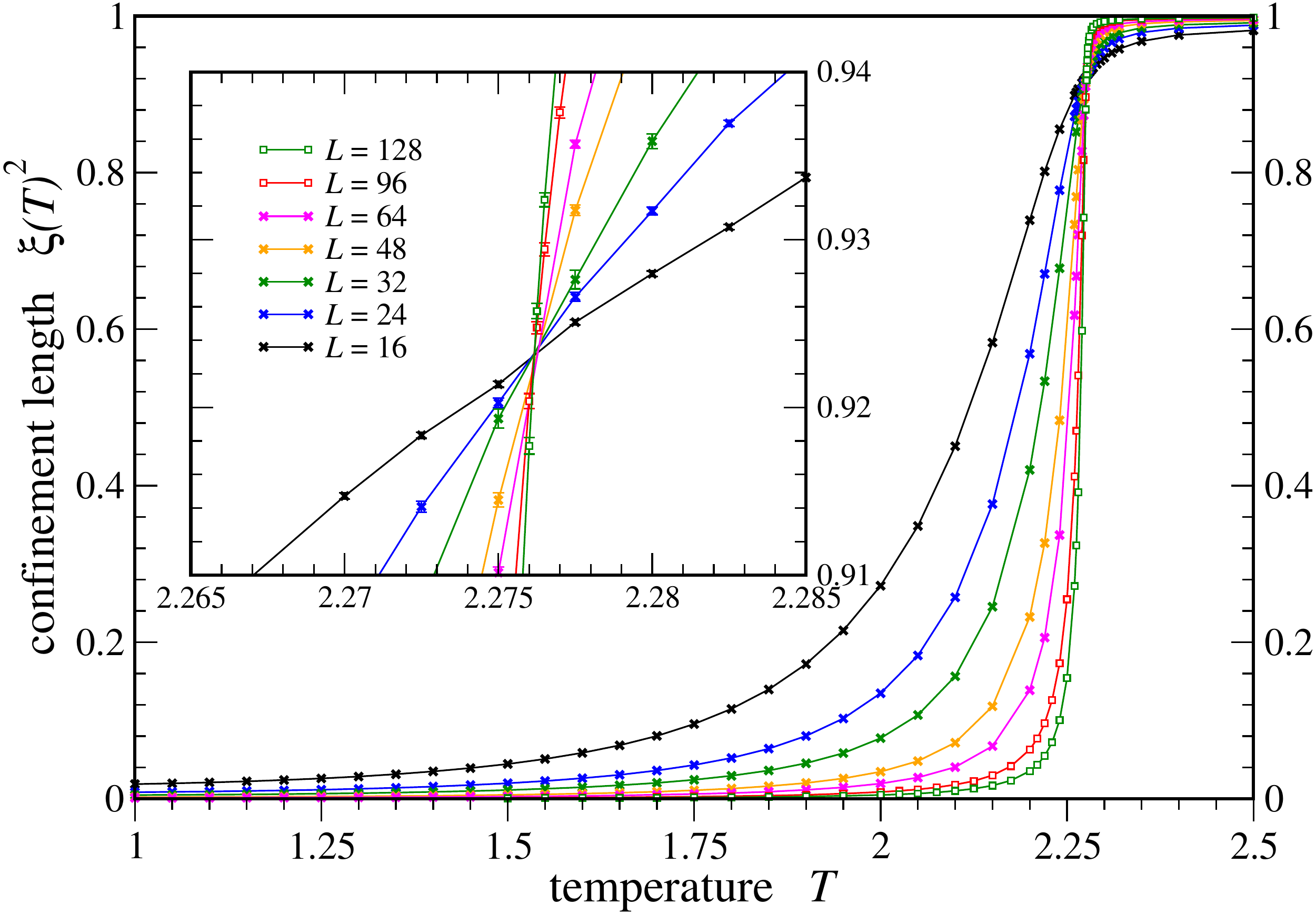}
\caption{
  (color online) 
  Confinement length $\xi^2(T)$ for the 1-GS model measuring the (squared) average distance
  between two monomers. 
  Data for different system sizes $L$ are renormalized by the expectation value $(L^2+2)/4$
  for deconfined monomers.
  The distinct crossing point shown in the inset for temperatures in the vicinity of the transition 
  temperature indicates a continuous transition.
   }
\label{Fig:1GS-Confinement}
\end{figure}

A distinct feature of the Coulomb phase is that (test) monomers are deconfined. 
As a consequence, we expect the monomers to confine at the phase transition out of the Coulomb
phase. This confinement transition can be tracked using the monomer confinement
length $\xi^2(T)$ introduced above. 
Plotting data for various systems sizes, as shown in Fig.~\ref{Fig:1GS-Confinement}, 
we observe a distinct crossing point at the transition temperature. 
This absence of finite-size effects at the transition temperature indicates a universal
value of the confinement length $\tilde{\xi}(T_c)$ at this transition,
which we estimate to be $\tilde{\xi}(T_c) \approx 0.923 \pm 0.001$.
This crossing point strongly indicates a continuous transition.

\begin{figure}[t]
\includegraphics[width=\columnwidth]{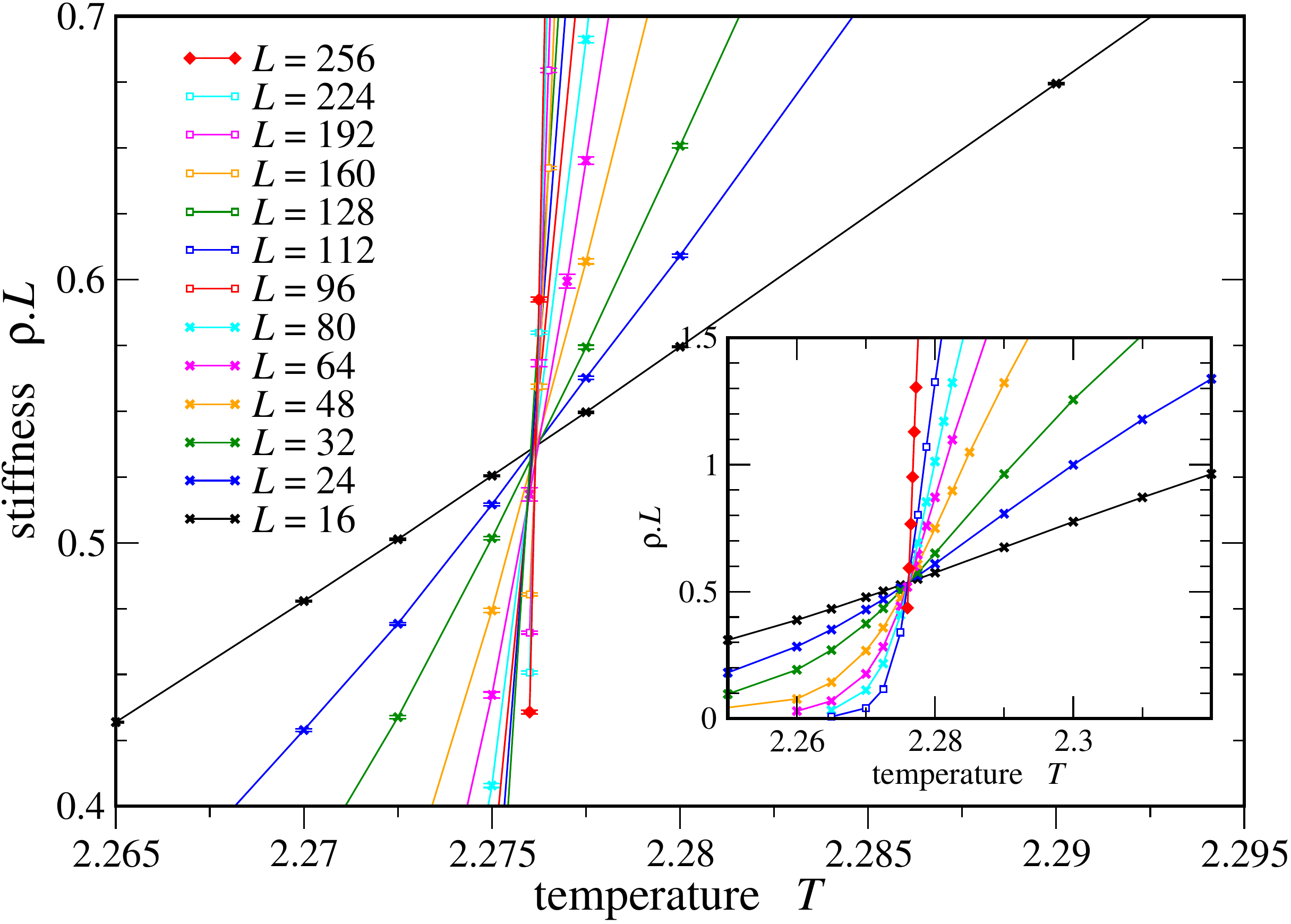}
\caption{
  (color online) 
  Stiffness $\rho$ of the 1-GS model multiplied by $L$ versus temperature $T$ near the transition, 
  for different system sizes $L$. 
  The crossing point indicates a continuous transition.
}
\label{Fig:1GS-Stiffness}
\end{figure}

Another indication of a continuous transition is that the distribution of dimer fluxes
$\phi$ also becomes universal at the transition temperature. Indeed we observe a distinct
crossing point for the stiffness $\rho$ (multiplied by system size $L$) when plotting curves for 
different $L$  in the vicinity of the transition out of the Coulomb phase, as shown in 
Fig.~\ref{Fig:1GS-Stiffness}.
The position of this crossing point coincides exactly with the transition temperature
$T_c =2.276 \pm 0.001$ estimated from the specific heat. 

Having established the continuous nature of the transition, we now turn to
its universality class. Since this phase transition occurs without any spontaneous 
symmetry breaking, we cannot rely on conventional techniques using an order
parameter to measure critical exponents. 
However, we can still consider thermodynamics, such as the behavior of the specific 
heat in the vicinity of the transition. As shown in Fig.~\ref{Fig:1GS-SpecHeat}, $C_v(T_c)/N$ 
grows very slowly with system size at criticality, which would suggest a critical exponent 
$\alpha>0$, but very small. 
It is also quite possible that $C_v(T_c)/N$ actually converges to a finite
value, but for system sizes that are currently out of reach of our numerical
simulations. This would indicate a negative value for  $\alpha<0$, also likely very small. 
This latter scenario is not unlikely considering the 3D XY model, which is known to have 
a small negative critical exponent $\alpha=-0.0151$~\cite{3DXY}, but for which numerical 
simulations~\cite{3DXYCV} do not see a convergence of the specific heat.

\begin{figure}[t]
\includegraphics[width=\columnwidth]{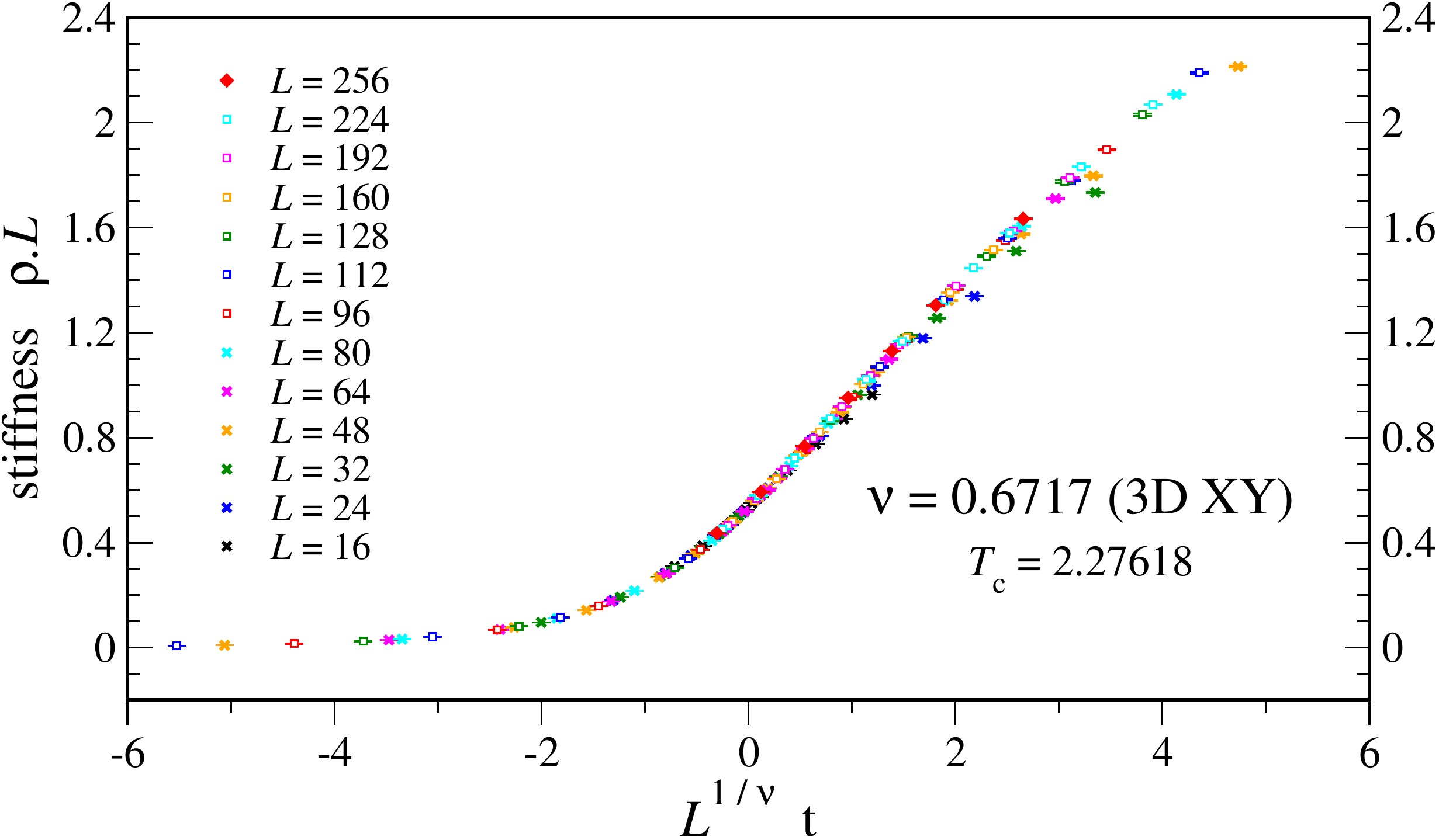}
\includegraphics[width=\columnwidth]{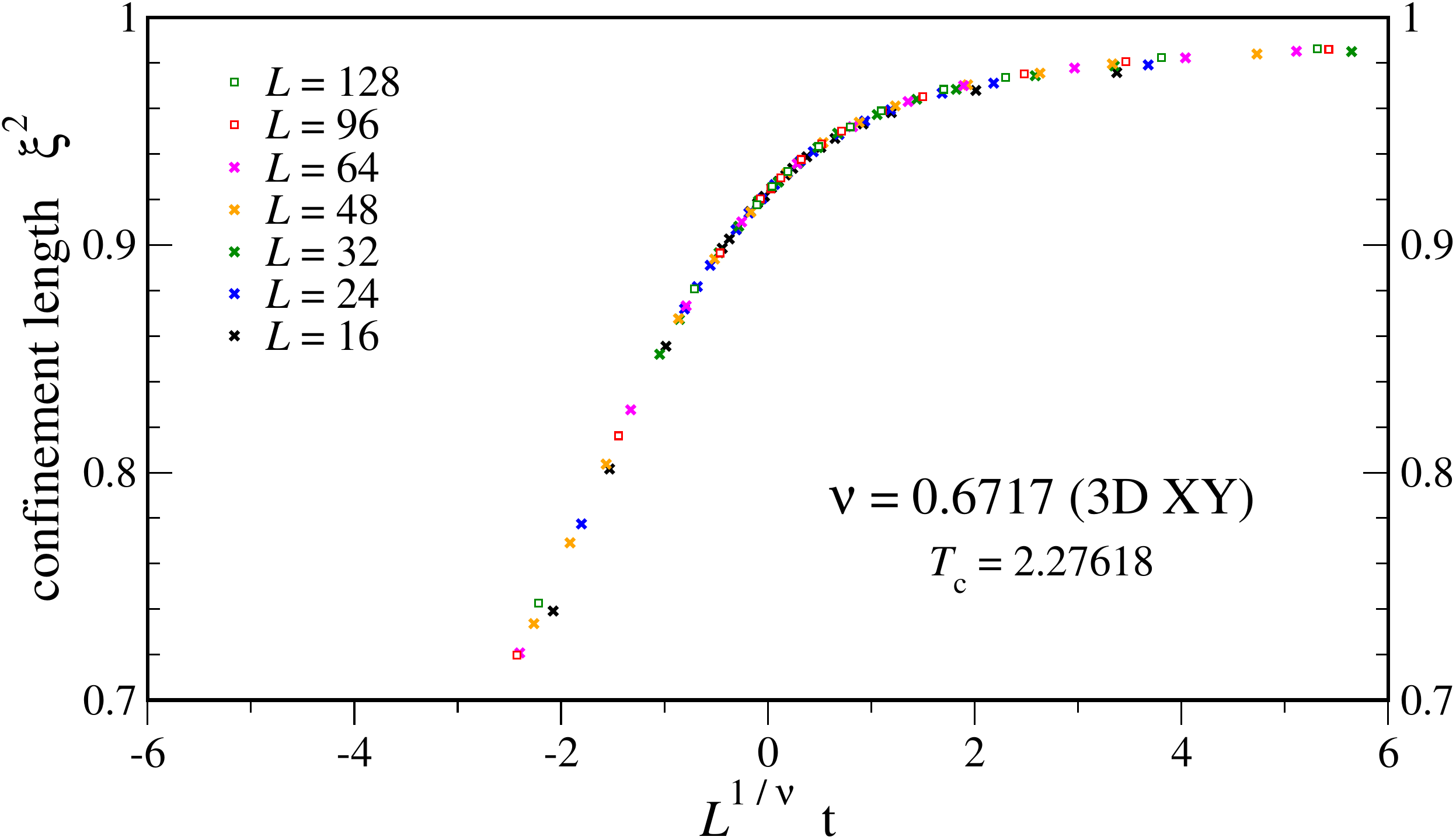}
\caption{
  (color online) 
  Data collapse for the stiffness  $\rho$ multiplied by $L$ (top panel) and confinement length 
  $\xi^2$ (lower panel) of the 1-GS model as a function of $L^{1/\nu} t$, 
  where $t = (T-T_c)/T_c$ with $T_c = 2.27618$.
  The critical exponent $\nu = 0.6717$ corresponds to the 3D XY universality class.
}
\label{Fig:1GS-DataCollapse}
\end{figure}

Thermodynamics being of little help to determine the universality class, 
another possibility is to consider crossings and data collapse of adequate quantities, 
including the stiffness and the confinement length. 
Standard finite-size scaling arguments indicate that close to the transition point, 
the stiffness should scale as $\rho=\frac{1}{L}\tilde{\rho}(L^{1/\nu} \cdot t)$, 
where $\tilde{\rho}$ is a universal function, $t=(T-T_c)/T_c$ the deviation from 
the critical temperature, and $\nu$ the correlation length exponent.
Performing this analysis, we find a nice data collapse  for the correlation length exponent 
$\nu = 0.6717$ of the 3D XY universality class~\cite{3DXY}
as shown in the top panel of Fig.~\ref{Fig:1GS-DataCollapse}.
The same scaling form $\xi = \tilde{\xi}(L^{1/\nu} \cdot t)$ is
also expected for the  confinement length $\xi^2$. 
As shown in the lower panel of Fig.~\ref{Fig:1GS-DataCollapse} we again find a data
collapse for the same exponent $\nu = 0.6717$.
Finally, we note that the system-size independent value $\tilde{\xi}(T_c) \approx 0.923 \pm 0.001$ 
is another characteristic of the universality class of the transition and in this case also points 
to the 3D XY universality class~\cite{note.link}.

\subsection{The 2-GS and 4-GS models}

\begin{figure}[t]
\includegraphics[width=\columnwidth]{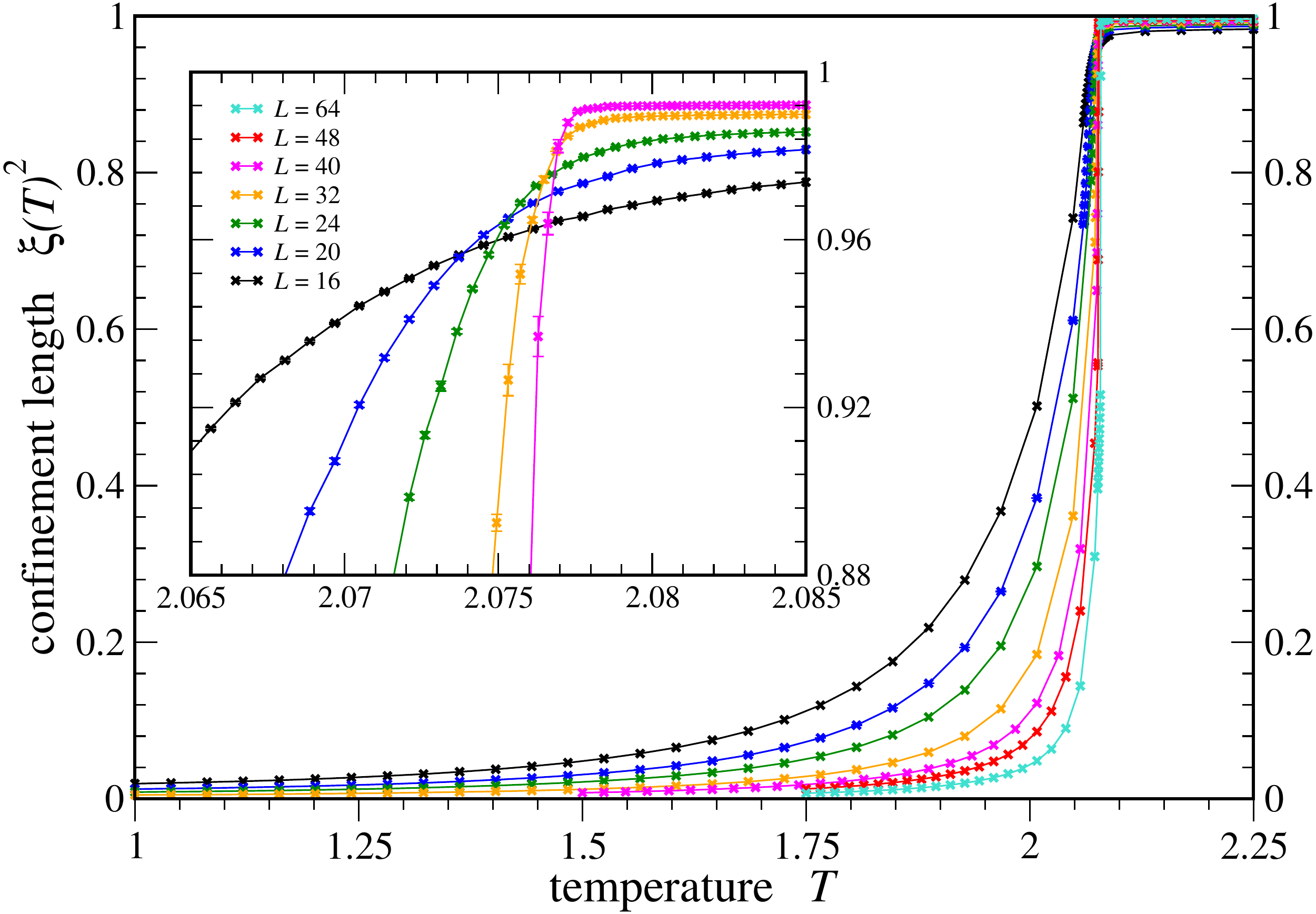}
\caption{
  (color online) Confinement length $\xi^2(T)$ for the 2-GS model.
  Data for different system sizes $L$ are renormalized by the expectation value $(L^2+2)/4$
  for deconfined monomers.
  The absence of a distinct crossing point in the vicinity of the transition temperature indicates
  a first-order transition.
}
\label{Fig:2GS-Confinement}
\end{figure}

\begin{figure}[t]
\includegraphics[width=\columnwidth]{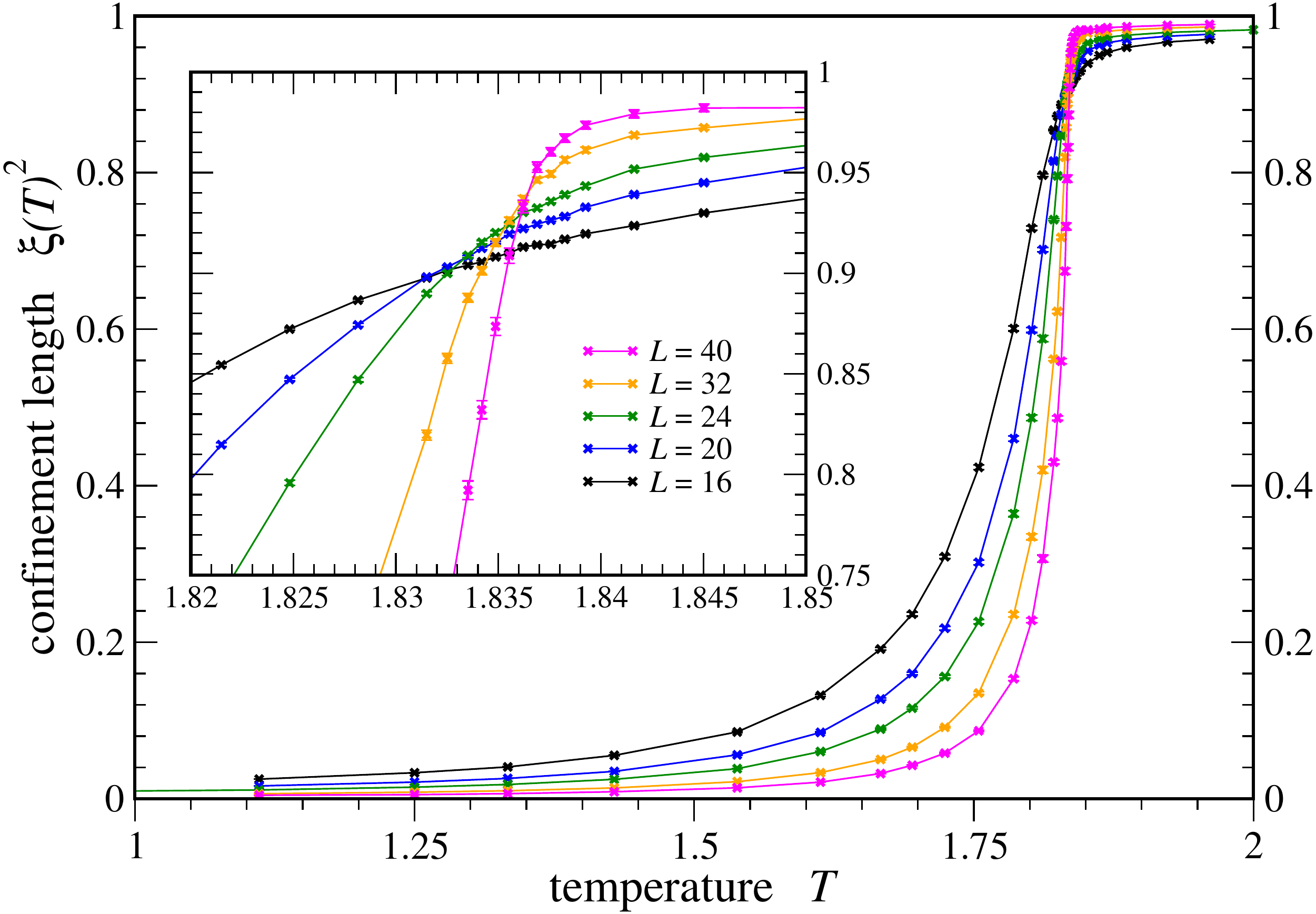}
\caption{
  (color online) Confinement length $\xi^2(T)$ for the 4-GS model.
  Data for different system sizes $L$ are renormalized by the expectation value $(L^2+2)/4$
  for deconfined monomers.
  The absence of a distinct crossing point in the vicinity of the transition temperature indicates
  a first-order transition.
}
\label{Fig:4GS-Confinement}
\end{figure}

We contrast our finding of a continuous transition in the 1-GS model with some numerical results
for the 2-GS and 4-GS models which both undergo first-order transitions between the Coulomb
phase and the dimer crystal phase. 
In Figs.~\ref{Fig:2GS-Confinement} and \ref{Fig:4GS-Confinement} the confinement length of
monomers is plotted. Similarly to the 1-GS model the transition out of the Coulomb phase is
accompanied with a confinement of the monomers. 
At these first-order transitions we do not observe a distinct crossing point, see the insets of 
Figs.~\ref{Fig:2GS-Confinement} and \ref{Fig:4GS-Confinement}.

\section{Continuous interpolation between dimer models}
\label{sec:interpolation}

One way to firmly establish the continuous nature of the phase transition in the 1-GS and 6-GS 
dimer  models is to demonstrate that these transitions can be embedded into {\em lines} of continuous transitions. We will first concentrate on the 1-GS model and show that such a line of continuous
transitions ending in a proposed, multicritical point can indeed be found. 
We will then discuss a similar idea for the
6-GS model, which however does not reveal such a line of continuous transitions.

\subsection{Interpolating the 1-GS and 2-GS models}
\label{sec:interpolation-1GS}

We have already observed that the 1-GS model which favors a single columnar ordering pattern in one lattice direction undergoes a continuous transition, while the 2-GS model which favors the two possible columnar ordering patterns in a given lattice direction undergoes a strong first-order transition. 
We can now ask how the nature of the phase transition changes as we continuously interpolate 
between these two models. 
To this end we continuously vary the weights for the columnar ordering patterns on the odd/even 
bonds in a given lattice direction. Formally, we introduce a coupling parameter $\lambda$ with
$0 \le \lambda \le 1$ on every other bond
\be
{\mathcal H}_{\rm 1-2-GS} = - \sum_{\substack{\square}} (\lambda n^o_{||} + n^e_{||})
\;.
\ee
For $\lambda=0$ we recover the 1-GS model, while $\lambda=1$ corresponds to the 2-GS model.

\begin{figure}
\includegraphics[width=\columnwidth]{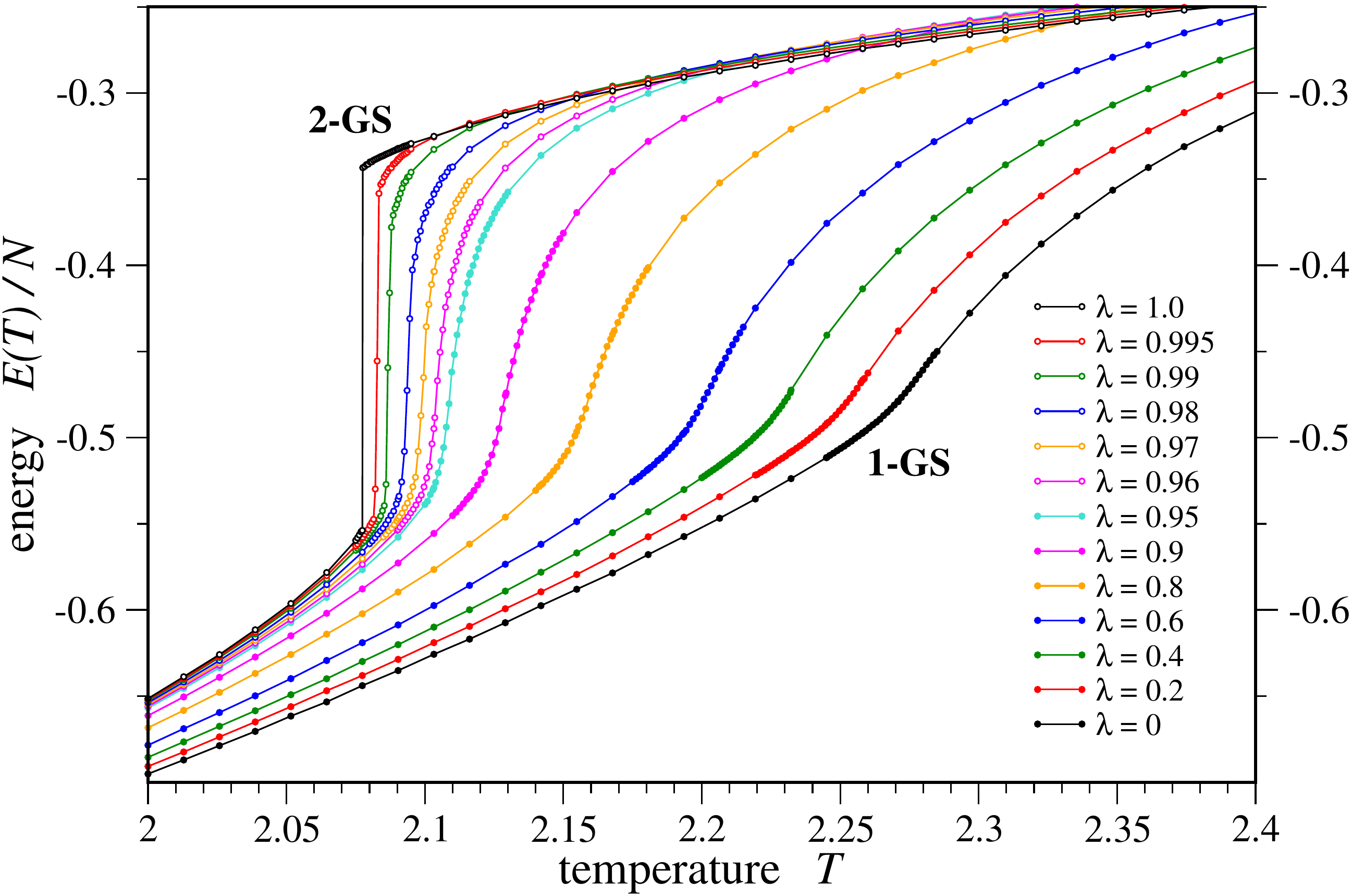}
\caption{
  (color online) 
  Energies per site for a dimer model which continuously interpolates
  between the 1-GS ($\lambda=0$) and 2-GS ($\lambda=1$) models.
  Data shown is for system size $L=48$.
}
\label{Fig:1GS-to-2GS-Energies}
\end{figure}

\begin{figure}
\includegraphics[width=\columnwidth]{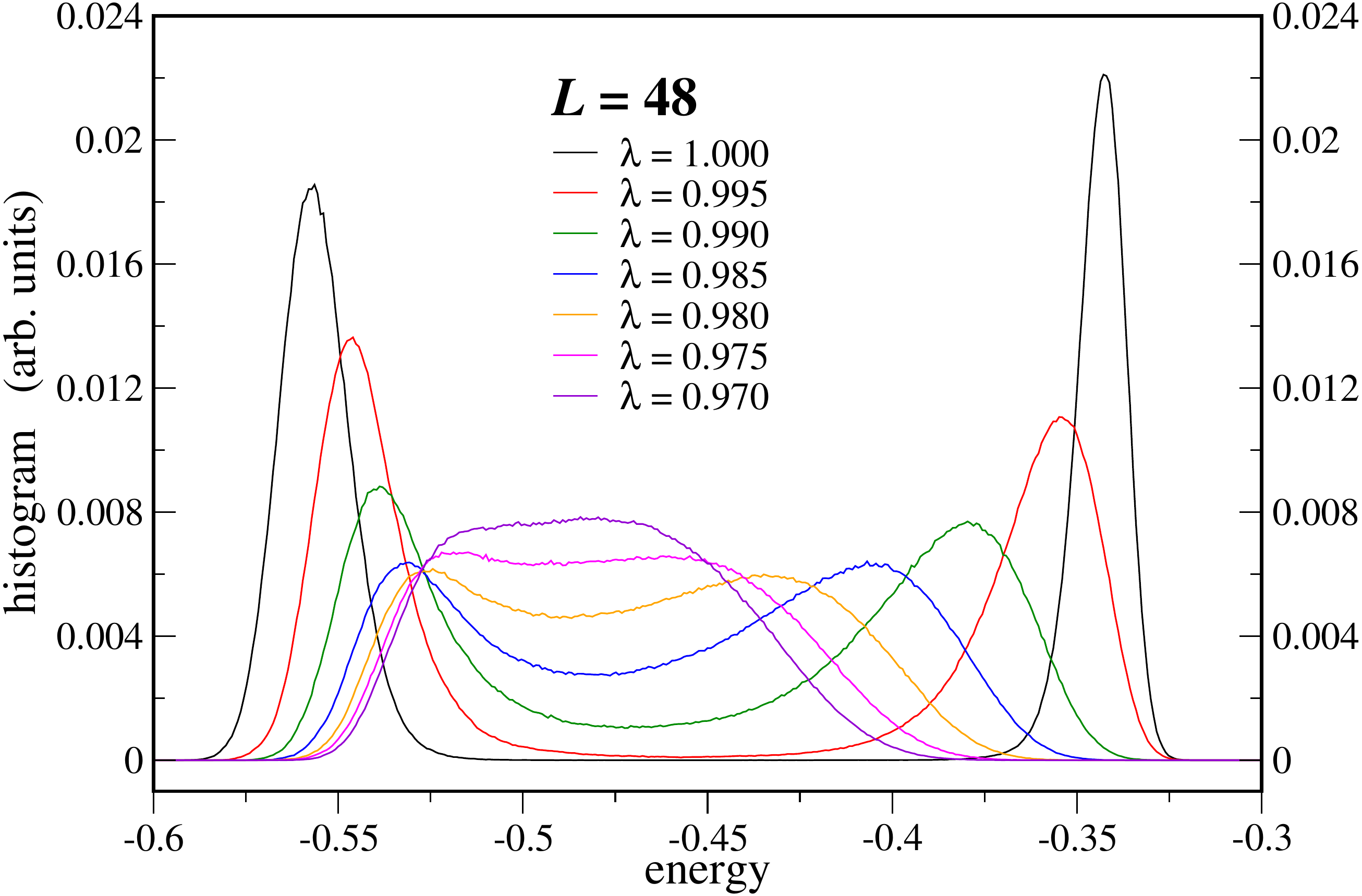}
\caption{
  (color online) 
  Energy (per site) histograms for the dimer model interpolating between the 1-GS and 
  2-GS models in the vicinity of the transition temperature of the respective
  models. Data shown is for system size $L=48$.
}
\label{Fig:1GS-to-2GS-Histograms}
\end{figure}

Our numerical simulations for various interpolation parameters $\lambda$ are summarized in 
Figs.~\ref{Fig:1GS-to-2GS-Energies} and \ref{Fig:1GS-to-2GS-Histograms}, which show the
energy per site $E(T)/N$ and histograms of the energy per site in the vicinity of the transition temperature, respectively.
Starting from the 2-GS model ($\lambda=1$) the sharp, kink-like feature in
the energy accompanying the first-order transition quickly vanishes for interpolation parameters $\lambda<1$ 
as the two possible columnar dimer orderings acquire different weights. 
The energy histograms in the vicinity of the transition temperatures  
turn from a bimodal distribution in the parameter regime $1 \ge \lambda \gtrsim 0.97$ into a single peak 
distribution for $\lambda < 0.97$ and system size $L=48$, see Fig.~\ref{Fig:1GS-to-2GS-Histograms}. 
This strongly suggests that the first-order transition of the 2-GS model turns
continuous for some intermediate $\lambda$, which for larger system sizes might be closer to 
$\lambda_c \approx 0.95$. On the other hand, this demonstrates that the continuous transition of the 1-GS model is indeed 
part of a line of continuous transitions which extends over the range 
$0 \le \lambda \le \lambda_c$ and likely ends in a {\em multicritical} point at $\lambda_c$.
Note that while $\lambda \neq 1$ introduces a staggering with respect to the two 
columnar ordering patterns, the system exhibits identical symmetries for all $0 \le \lambda < 1$,
which will lead to a uniform theoretical description of the interpolated models for all $0 \le \lambda < 1$. 
However, since the strong first order transition of the 2-GS is expected to
be stable towards a small perturbation $\delta \lambda = 1-\lambda$ it is not surprising to see that the interpolated models 
exhibit (weak) first-order transitions in the regime $0.97 \lesssim \lambda \le 1$, but quickly turn to uniform behavior
for smaller $\lambda$.

\begin{figure}[htp]
\includegraphics[width=\columnwidth]{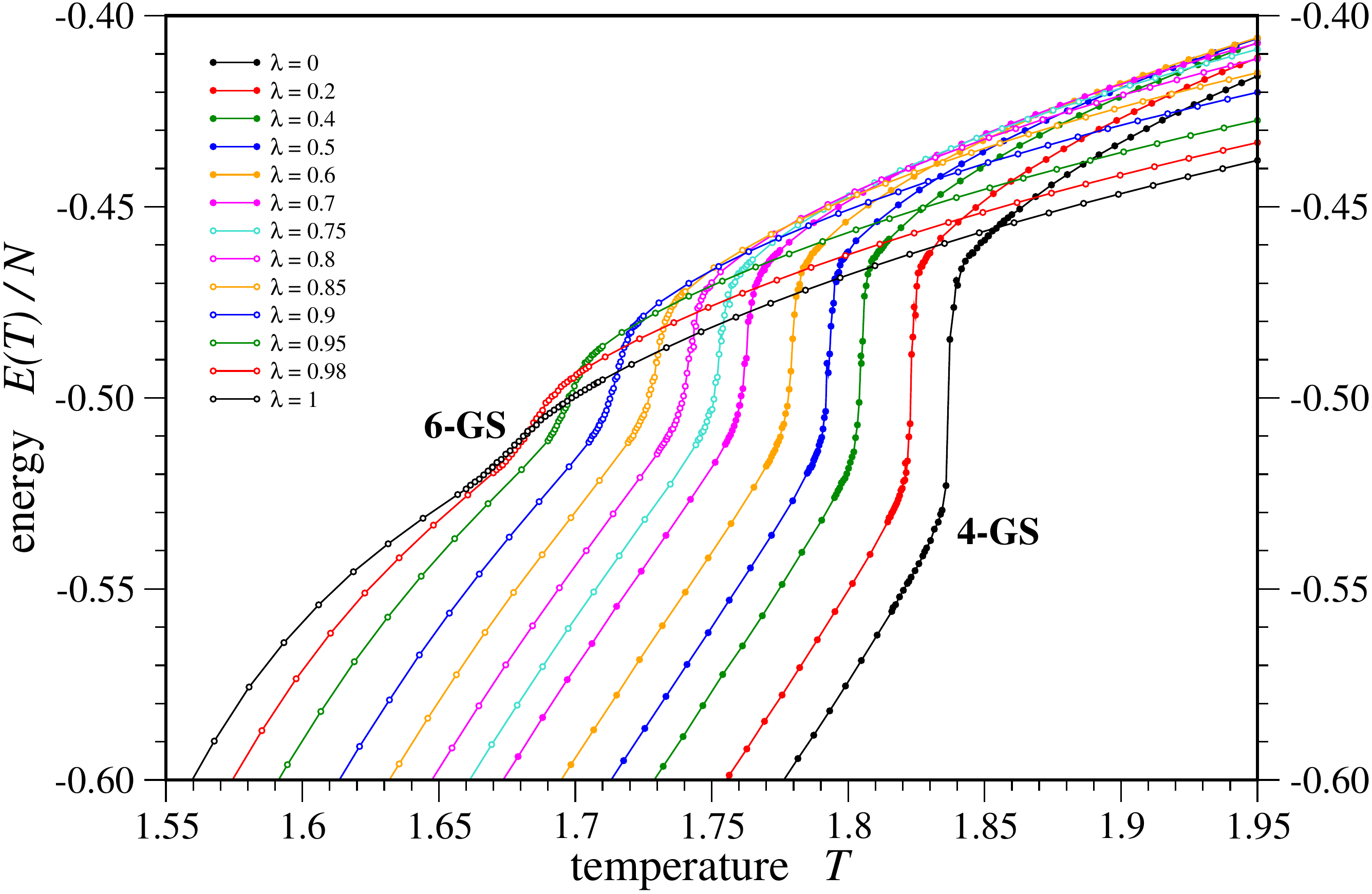}
\caption{
  (color online) 
  Energies per site for a dimer model which continuously interpolates
  between the 4-GS ($\lambda=0$) and 6-GS ($\lambda=1$) models.
 Data shown is for system size $L=48$.
}
\label{Fig:4GS-to-6GS-Energies}
\end{figure}

\begin{figure}[htp]
\includegraphics[width=\columnwidth]{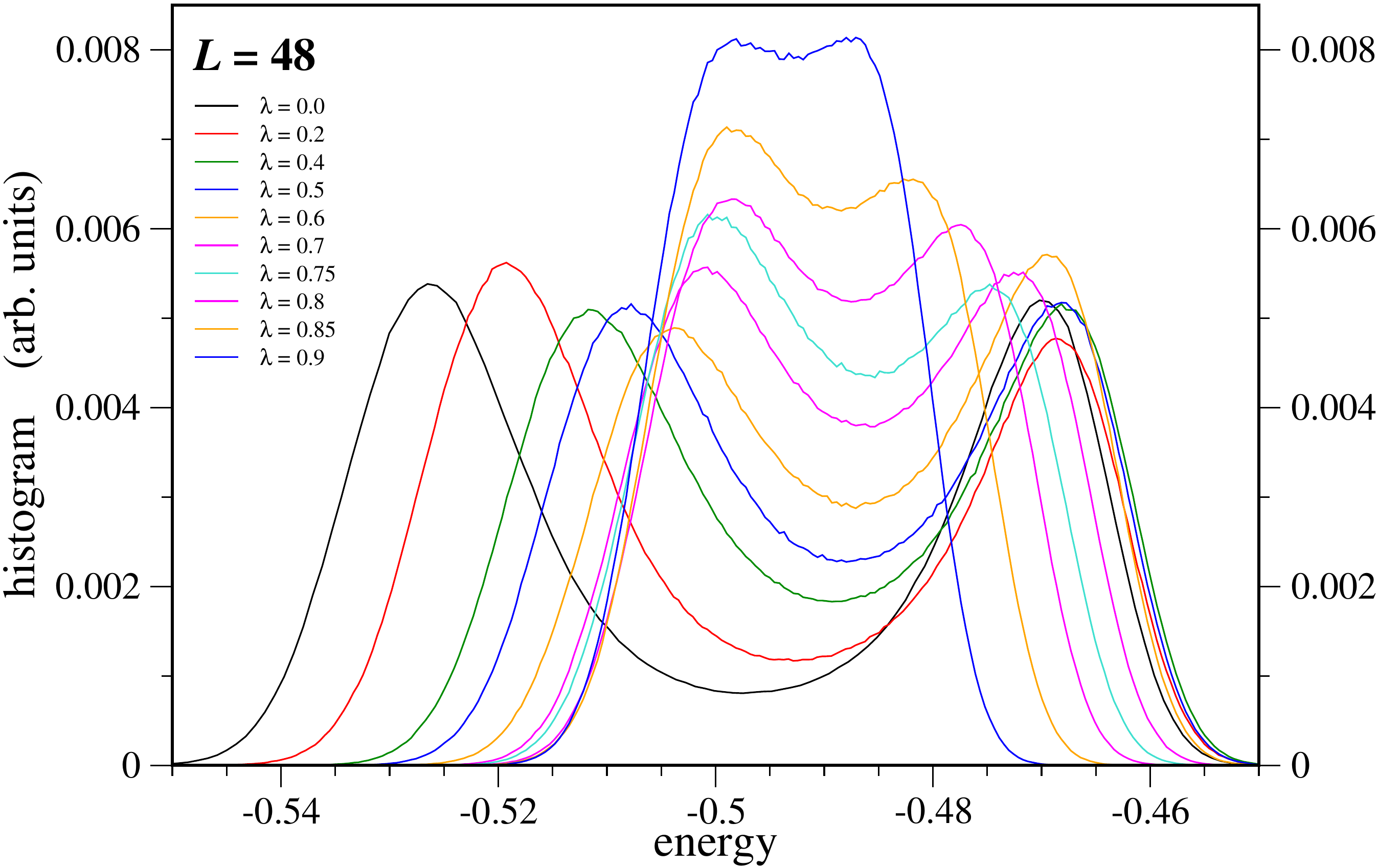}
\vskip 4mm
\includegraphics[width=\columnwidth]{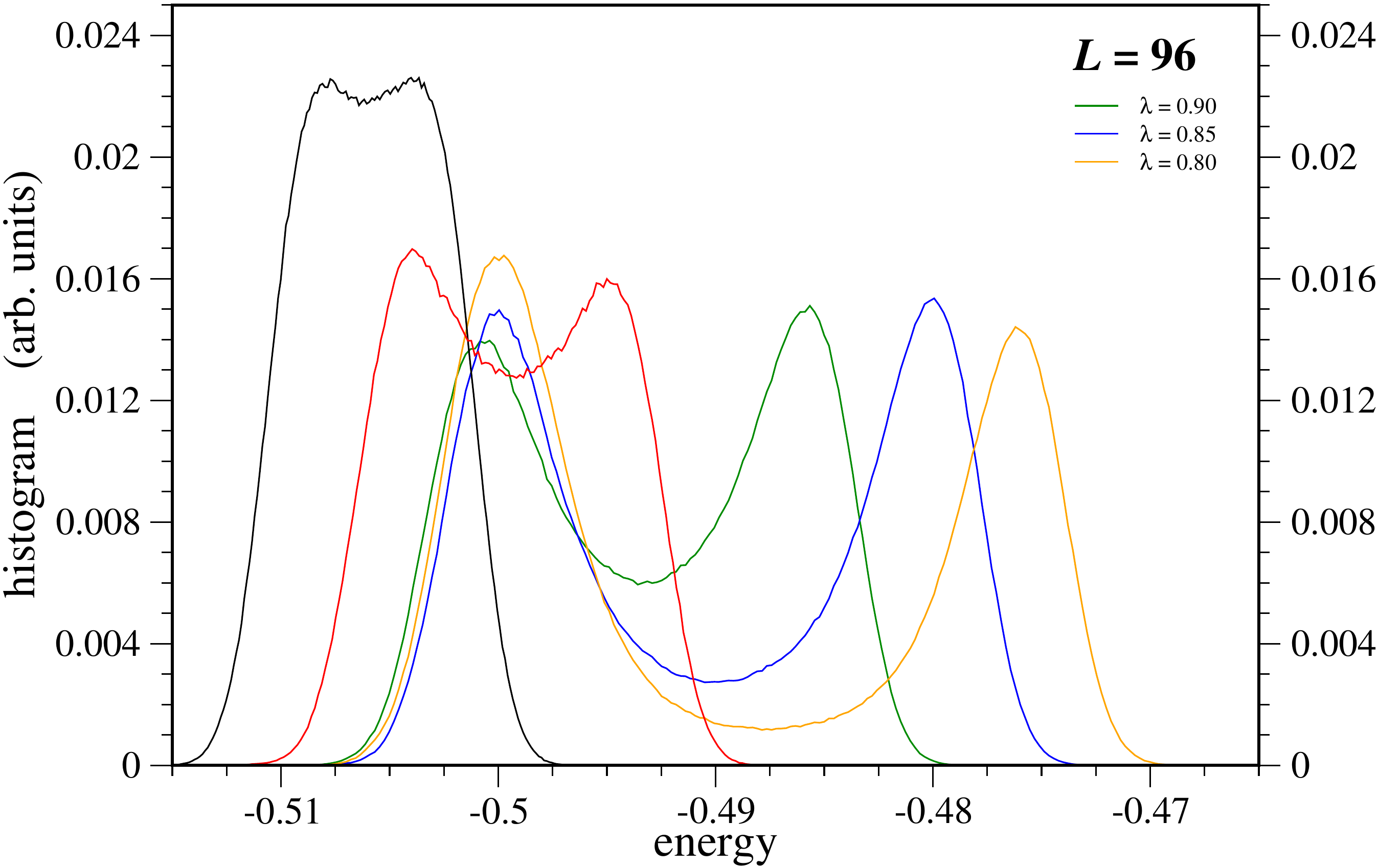}
\caption{
  (color online) Energy (per site) histograms for the dimer model interpolating between the 4-GS to 6-GS models
  in the vicinity of the transition temperature of the respective models. Data shown are for system size
  $L=48$ in the upper panel and $L=96$ in the lower panel.
}
\label{Fig:4GS-to-6GS-Histograms}
\end{figure}

\begin{figure}[htbp]
\includegraphics[width=0.95\columnwidth]{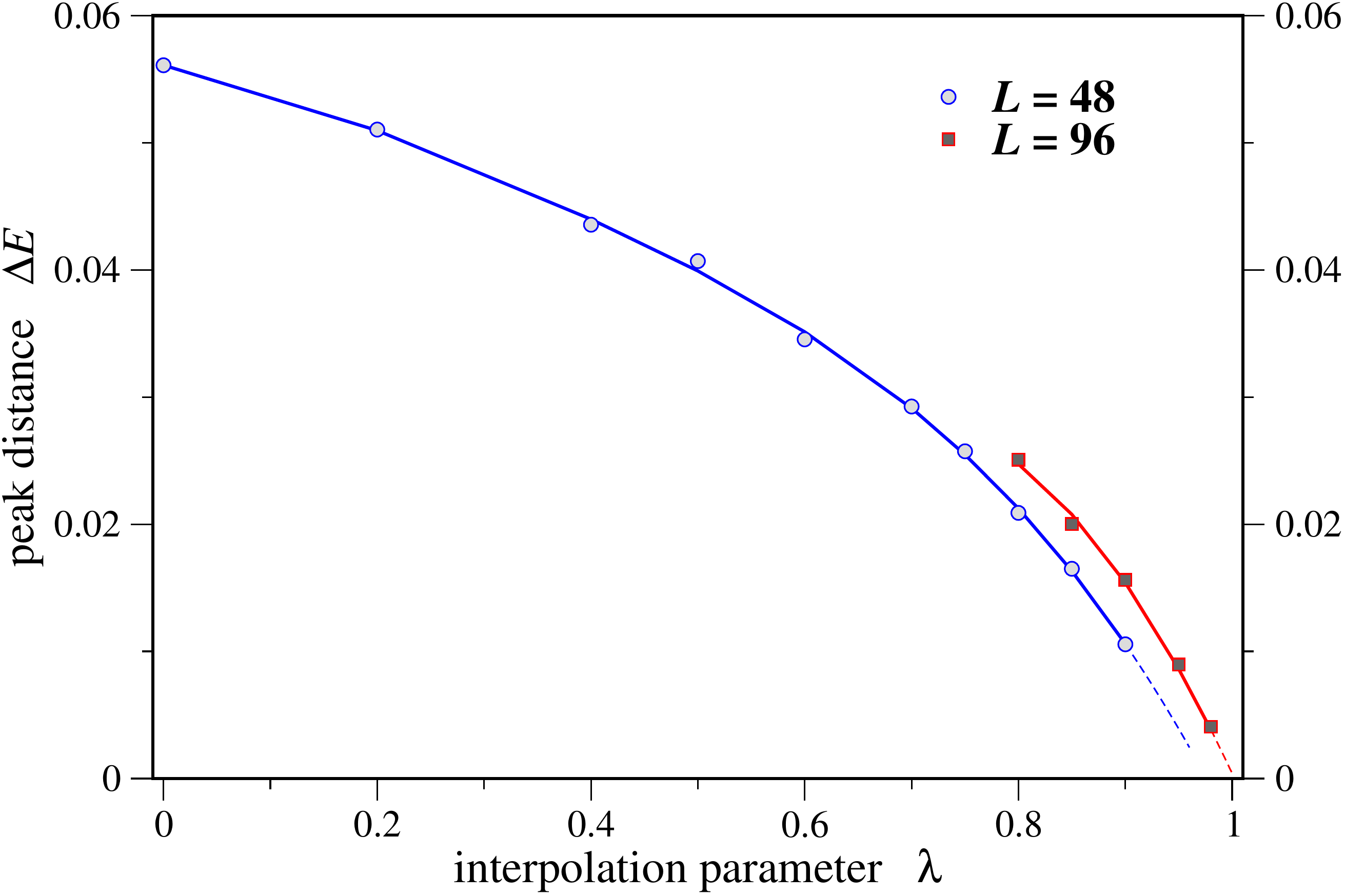}
\caption{
  (color online) Distance between the two peaks in the bimodal energy histograms  
  shown in Fig.~\ref{Fig:4GS-to-6GS-Histograms}. The lines are guides to the eye.
  Data for system sizes $L=48$ and $L=96$ are shown.
}
\label{Fig:4GS-to-6GS-HistogramPeaks}
\end{figure}

\subsection{Interpolating the 4-GS and 6-GS models}
\label{sec:interpolation-6GS}

We now turn to the 6-GS model, for which we can study a similar interpolation to the 4-GS model,
which in contrast to the 6-GS model undergoes a first-order transition.
Again we introduce an interpolation parameter $\lambda$ which now assigns different weights to
the two columnar ordering patterns that establish the difference between the 4-GS and 6-GS model.
Formally, we investigate the Hamiltonian
\be
{\mathcal H}_{\rm 4-6-GS} = - \sum_{\substack{\square}} (n_{=} + n_{//} + \lambda n_{||})
\;,
\ee
with $0 \le \lambda \le 1$.
For $\lambda=0$ this is the 4-GS model, while $\lambda=1$ now corresponds to the 6-GS model.

Our numerical results for various interpolation parameters $\lambda$ are summarized in 
Figs.~\ref{Fig:4GS-to-6GS-Energies}, \ref{Fig:4GS-to-6GS-Histograms} and
\ref{Fig:4GS-to-6GS-HistogramPeaks}, respectively.
We find that a sharp, kink-like feature around the transition temperature in the energy $E(T)$ persists
for almost all interpolation parameters $0 \le \lambda < 1$ (see
Fig.~\ref{Fig:4GS-to-6GS-Energies}). Energy histograms in Fig.~\ref{Fig:4GS-to-6GS-Histograms} for the respective
transition temperatures show bimodal distributions for the same parameter
range. Pushing the limit of our calculations we can establish a two-peak structure up to $\lambda = 0.98$ for system size 
$L=96$ (see the lower panel in Fig.~\ref{Fig:4GS-to-6GS-Histograms}). 
If we systematically trace the distance between the two peaks in these bimodal energy distributions,
as shown in Fig.~\ref{Fig:4GS-to-6GS-HistogramPeaks}, the emerging trend clearly suggests that the
line of first-order transitions persists all the way up to $\lambda < 1$, but the transition turns weaker 
along this line with the data suggesting that there is no bimodal distribution for $\lambda=1$ and the 
transition becomes continuous for the 6-GS model.


\section{Candidate Field Theories}
\label{sec:field}

To understand the different nature of the phase transitions discussed in
Sec.~\ref{sec:numerics}, we will now develop a family of candidate field
theories describing these transitions.   To do this, we take advantage
of a pair of mappings: from the dimer model to compact lattice quantum
electrodynamics (QED), and thence to a dual monopole formulation.  The
continuum limit of the latter leads directly to the desired field
theories.  

The desired mappings have been discussed in some detail in
Refs.\onlinecite{senthil:prb2005},\onlinecite{bergman:prb2006}, and in
fact can be performed not only for the classical dimer model discussed
here but also its quantum generalization.  For clarity, we will present
in this section a brief, self-contained summary of the mapping in the
classical case for the cubic lattice, applicable to the models of the
present paper.  Performing a detailed symmetry
analysis, we then derive the symmetry allowed Ginzburg-Landau actions for
the various models and discuss implications for their respective phase
transitions.

\subsection{Mappings}
\label{sec:mappings}

To proceed, we define a bond variable which counts the number of dimers
on a given bond: 
\begin{equation}
\hat{n}_{ab} = 
\left\{
    \begin{array}{lc} 
      1\,, & \textrm{if the bond  $\langle a,b\rangle $ is occupied by a dimer} \\ 
      0 & \textrm{otherwise.} 
    \end{array}
\right.
\label{eq:ndef}
\end{equation}
The close-packed dimer constraint which requires that every site in the cubic lattice
be part  of exactly one dimer can then be expressed as
\be
\label{1boson}
\hat N_a \equiv \sum_{b} {\hat n}_{ab} = 1\;, \ee 
where the sum is over
sites $b$ which are nearest neighbors of $a$.  A monomer excitation in
the dimer model which breaks the close-packing constraint, e.g. an
unpaired site $a$ on the cubic lattice, is then indicated by $\hat N_a =
0$.

\subsubsection{compact QED}
\label{sec:compact-qed}

We may directly pass to QED variables as follows. 
We introduce an electric field variable $E_{a b}$, which is a directed
variable, according to
\begin{align}
E_{a b} = \epsilon_a \hat{n}_{a b}\;, 
\label{eq:E_ab}
\end{align}
where $E_{a b}$ is integer-valued (in particular $E_{a b}=0,\pm 1$) and
we have introduced a `background charge' $\epsilon_a$ with a fixed distribution
of alternating charges on the two sublattices
\be
\epsilon_a = 
\left\{
    \begin{array}{ll}
      +1 & , \, a \in \textrm{A sublattice,}   \\
      -1 & , \, a \in \textrm{B sublattice.}
\end{array}
\right.
\label{eq:e_a}
\ee

In the QED formulation, the local constraint (\ref{1boson}) maps directly to
a lattice version of the Gauss law,
\be
\label{Gauss} 
\divo \vec E=\epsilon_a \;,
\ee
which also explains the notion of the background charge and
where we have used the lattice divergence $ \divo \vec E = \sum_b E_{a b} $. 

Expressed in the QED variables, the Hamiltonian becomes
\bea
{\mathcal H}_{\rm QED} & = & \frac{U}{2} \sum_{\substack{\langle a,b\rangle }} 
( E_{a b} - \frac{\epsilon_a}{2}  )^{2}
-  \sum_{\substack{\square}} (\delta_{\curl {\vec E}, 2} +
\delta_{\curl {\vec E}, -2} )  \nonumber \\
& & + \textrm{ const.} \,
\label{QED1}
\eea
where the first term is a constant in the physical space in which
$\hat{n}_{ab}=0,1$.  We include it, however, in order that we may allow
the electric variable $E_{ab}$ to fluctuate over {\sl all} integers; by
taking the large $U$ limit, the physical dimer states, which minimize
this term, are selected.  It is expected that the universal properties
of ${\mathcal H}_{\rm QED}$ are identical for infinite and finite $U$.  

Note that the Hamiltonian (\ref{QED1}) is rather similar to the standard
formulation of compact QED. The main difference is the absence of any
magnetic field terms $B^2$, which reflects the classical nature of the
dimer model under consideration, a shifting of the $E^2$ term by an
alternating `background field' $\epsilon_a/2$ and the energetic
preference for
$\curl {\vec E}=\pm 2$.
Despite these differences, Hamiltonian (\ref{QED1}) does
share all the same internal symmetries as the more conventional QED
form. It is therefore expected to share the same properties in
regimes where universality is mandated.

\subsubsection{Duality, Monopole Formulation and Symmetry Transformation}
\label{sec:duality}

In contrast to a conventional, non-compact QED formulation a {\em
  compact} QED like the one introduced in the previous section does {\em
  not} prohibit magnetic monopoles.  These monopoles are `conjugate' to
the gauge charges in the QED description, which in terms of the original
dimer models correspond to monomer excitations.  The defining
characteristic of the Coulomb phase is that the gauge charges are {\em
  deconfined}, while the magnetic monopoles are well-defined, gapped
quasiparticles.  In the dimer crystal phase, on the other hand, the
electric charges are confined and the electric field is static. This
implies that the conjugate magnetic field is strongly fluctuating and
the magnetic monopoles are no longer good excitations.
We can thus describe the phase transition out of the Coulomb phase as
the (Bose) condensation of magnetic monopoles accompanied by a
simultaneous confinement transition of the electric charges.

Our goal now is to establish an analytic description of this phase
transition in terms of the magnetic monopoles. To this end, we will
first introduce a duality transformation to make the monopole
excitations explicit in the Coulomb phase.  It is well-known that the
electric and magnetic fields in Maxwell's equations are dual in the {\em
  absence} of charges and currents.
However, while there are no currents in our system, there is a
non-vanishing charge distribution, as given by the non-vanishing
electric field divergence in Eq.~\eqref{Gauss}.
We therefore introduce a `background electric field' ${\vec e}^{\,(0)}$,
which compensates for these background charges by satisfying
\be 
	\divo {\vec e} = \epsilon_a,
	\label{Eq:background-field}
\ee 
and making $\vec{E} -  {\vec e}$ divergence free.
Note that while $\vec{E}$ is a fluctuating variable, the background
field ${\vec e}$ is static.  
It is convenient to choose $\vec{e}$ to be integer valued, for what
follows.  Then a simple choice to satisfy (\ref{Eq:background-field})
is to place $e_{a,a+\hat{x}} = (1+\epsilon_a)/2 $ on the link emanating
in the $x$-direction from site $a$.

This allows us to write down a duality transformation of the form
\begin{equation}
{\vec E}_{ab}  = -\curl {\vec \alpha} + {\vec e}_{ab} \;,
\end{equation}
where we have introduced a  dual electric vector potential ${\vec{\alpha}}$.
In a quantum theory,  ${\vec{\alpha}}$ would generate
the conjugate magnetic field.
Due to the integer constraint on $\vec{E}_{ab}$, we must also take $\vec
\alpha$ to be integer valued (this reflects our integer choice of
$\vec{e}_{ab}$).  

\begin{figure}[t]
\includegraphics[width=0.8\columnwidth]{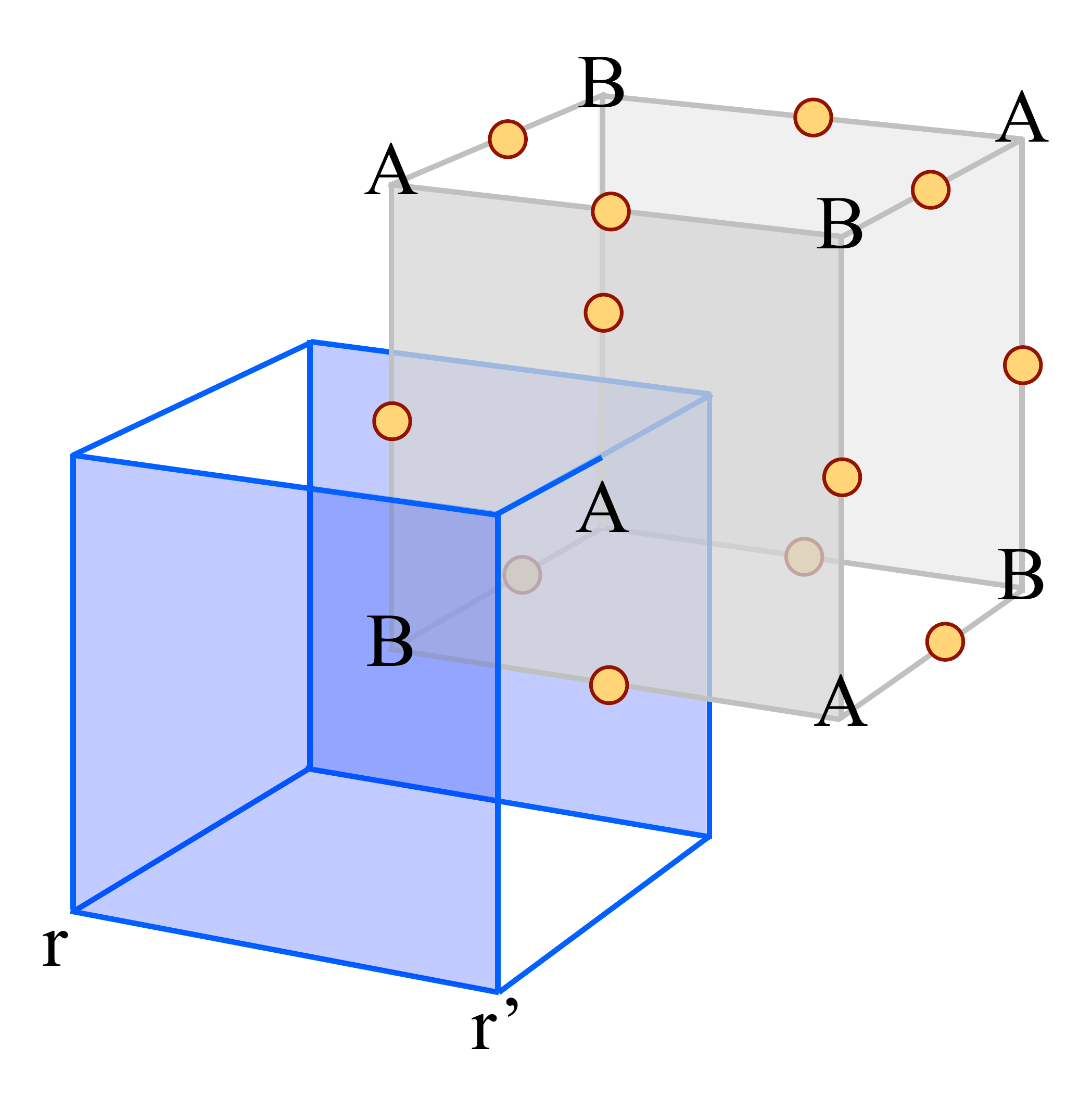}
\caption{(color online). Cubic lattice and its dual. The dual cubic lattice sites sit
at the centers of each cube of the cubic lattice.
}
\label{Fig:duallattice}
\end{figure}

In terms of these dual variables, the QED Hamiltonian (\ref{QED1}) becomes
\begin{eqnarray}
{\mathcal H}_{\rm dual-QED} & = &  \frac{U}{2} \sum_{\substack{\square}}
\left( \curl {\vec \alpha} - \vec{e}  + \frac{\epsilon_a}{2}  \right)^{2} \nonumber \\
&& -  \sum_{\langle ab\rangle}  F[\nabla^2 \alpha] \;,
\label{dual_QED}
\end{eqnarray}
where the squares `$\square$' now denote the plaquettes of the dual cubic lattice,
see Fig.~\ref{Fig:duallattice}, and the last term, $F[\nabla^2\alpha]$
represents the transcription of the last term in Eq.~\eqref{QED1} in terms
of $\vec\alpha$.  We will not need its explicit form here.  It is only
important to note that it contains {\sl second order} (lattice)
derivatives of the vector potential $\vec\alpha$.  Upon coarse-graining, such terms are
irrelevant in the continuum limit.  What is important is that, in the
process of integrating out short scale fluctuations, they will generate
relevant terms of all possible types dictated by symmetry.  In this way,
the physics of the dimer interactions, which is reflected in the $F$
function, enters the low energy continuum field theory description.

We now proceed to develop the continuum limit, following a sequence of
standard manipulations \cite{Kogut}.  In doing so, we will neglect the
$F$ term, on the grounds discussed above, keeping in mind that in the
final continuum theory, we must restore all possible symmetry-allowed
interactions that may be generated from it.  We first soften the integer
constraint on the vector potential ${\vec \alpha}$, replacing the
constraint by a term which favors integer values. This approximation
does not change the nature of the monopole condensing phase transition.
We rewrite the Hamiltonian (\ref{dual_QED}) as \be {\mathcal H} =
\frac{U}{2} \sum_{\substack{\square}} \left( \curl {\vec \alpha}
  -\vec{e}\right)^{2} - w \sum_{r,r'} \cos [2\pi \alpha_{r,r'}] \;,
\label{gaugedH}
\ee
where large $w$ recovers the integer $\alpha$ constraint.  With this
rewriting, we may regard $\alpha$ as a real-valued variable.  In the
first term we have also {\sl dropped} the $\epsilon_a/2$ term found
inside the parenthesis of the first term in Eq.\eqref{dual_QED}.  This is
possible because this term, regarded as a vector field, is purely
longitudinal (i.e. curl-free), and hence, actually decoupled from the
$\curl \vec\alpha$ factor.  We will soon extract a further longitudinal
piece from $e$.

To proceed, we first introduce explicit monopole phase variables by
making the gauge transformation $\alpha_{rr'} \rightarrow \alpha_{rr'} -
(\theta_r - \theta_{r'})/2\pi$.  One obtains
\begin{equation}
  \label{eq:1}
  {\mathcal H} = \frac{U}{2}
\sum_{\substack{\square}} \left( \curl {\vec \alpha}
  -\vec{e}\right)^{2} - w \sum_{r,r'} \cos [\theta_r-\theta_{r'} -
2\pi \alpha_{r,r'}] \;.
\end{equation}

Next, we break the background field $e$ into transverse and
longitudinal parts,
\begin{equation}
  \label{eq:2}
  \vec{e} = \vec{e}_L + \vec{e}_T,
\end{equation}
such that $\divo \vec e_T=0$ and ${\rm curl}\, \vec e_L = 0$.  Note that, because of
the $\curl$ in Eq.~\eqref{eq:1}, only $\vec e_T$ couples to $\vec \alpha$.  Taking
the divergence of Eq.~\eqref{eq:2}, we see that $\divo \vec e_L = \divo
\vec e =
\epsilon_a$.  A choice for $\vec e_L$ satisfying this condition {\sl and}
which is curl-free is simply
\begin{equation}
  \label{eq:4}
  (e_L)_{a,a+\mu} = \epsilon_a/6. 
\end{equation}
From this, we of course can find $e_T$ by solving Eq.~\eqref{eq:2}:
\begin{equation}
  \label{eq:6}
  (e_T)_{a,a+\mu} =\frac{1}{2}(1+\epsilon_a)\delta_{\mu,x} - \epsilon_a/6 \; .
\end{equation}
Inserting Eq.~\eqref{eq:2} into Eq.~\eqref{eq:1}, and dropping the
decoupled and constant $e_L$ part, we find
\begin{equation}
  \label{eq:5}
   {\mathcal H} = \frac{U}{2}
\sum_{\substack{\square}} \left( \curl {\vec \alpha}
  - \vec{e}_T\right)^{2} - w \sum_{r,r'} \cos [\theta_r-\theta_{r'} -
2\pi \alpha_{r,r'}] \;.
\end{equation}

At this point, we have obtained a lattice Ginzburg-Landau theory, in
which $\vec e_T$ appears as the (average) dual flux (experienced by the
monopoles) through the dual plaquette pierced by this vector, expressed
in units of the flux quantum.  As usual, only the fractional part of the
flux has physical significance.  From Eq.~\eqref{eq:6}, we readily see
that this fractional part is uniformly $1/6$ of a full flux quantum
piercing the dual plaquettes emanating from one sublattice of the direct
lattice and ending in the other.  This can be seen as an array of
alternating dual monopole fluxes, representing the original alternating
$\epsilon_a$ background charges in the QED theory.

Precisely this problem, of monopoles moving in this background flux
pattern on the cubic lattice, was studied by Motrunich and Senthil, in
Ref.~\onlinecite{senthil:prb2005}.  We can adapt their results
directly.  We define a soft-spin monopole field $\psi_r \sim
e^{i\theta_r}$, and neglect at first the fluctuations in the gauge
field, replacing $\alpha_{rr'}$ by a static gauge configuration
$\overline\alpha_{rr'}$ representing the background flux.  The soft-spin
Hamiltonian is 
\be 
\label{H1} 
{\mathcal H}_{\rm monopoles} = - w \sum_{ \langle r , r'
 \rangle } \left[ {\psi^\dag}_{r'} \, \psi_r \, e^{-2\pi i
   \overline{\alpha}_{r,r'}} + {\rm h.c.}  \right] 
\; . 
\ee 

Next we take the continuum limit, following Sec.~VI.~B of
Ref.~\onlinecite{senthil:prb2005}.  The hopping Hamiltonian in
Eq.~\eqref{H1} has two minima, which we will call $\Psi_1(r)$ and $\Psi_2(r)$ here.
A solution of the tight-binding Hamiltonian then becomes a linear
combination 
\be
\psi(r) = \phi_1 \Psi_1(r) + \phi_2 \Psi_2(r) ,
\ee
where we treat $\phi_1$ and $\phi_2$ as slowly varying fields.

We can now ask how these solutions transform under the symmetry operations
of the original cubic lattice. For the specific gauge choice of Ref.~\onlinecite{senthil:prb2005}
these were reported to be 
\bea
\left\{
\begin{array}{lll}
T_x:&  \phi_1 \rightarrow \phi_1^{\ast},&  \phi_2 \rightarrow -\phi_2^{\ast}       \\
T_y:&  \phi_1 \rightarrow \phi_1^{\ast},&  \phi_2 \rightarrow  \phi_2^{\ast}       \\
T_z:&  \phi_1 \rightarrow \phi_2^{\ast},&  \phi_2 \rightarrow  \phi_1^{\ast}       \\
R_{\frac{\pi}{2},Rxy}: &  \phi_1 \rightarrow  e^{-i\pi/4}\phi_1^{\ast},
                      &  \phi_2  \rightarrow e^{i\pi/4} \phi_2^{\ast}             \\
R_{\frac{\pi}{2},Rxz}: &  \phi_1 \rightarrow  \frac{1}{\sqrt{2}}(\phi_1^{\ast}+\phi_2^{\ast}),
                      &  \phi_2 \rightarrow  \frac{1}{\sqrt{2}} (\phi_1^{\ast}-\phi_2^{\ast}) \;,
\end{array}
\right.
\label{eq:sym-phi}
\eea
where the two $90^{\circ}$ rotations $R_{\frac{\pi}{2},Rxy}$ and $R_{\frac{\pi}{2},Rxz}$ 
along the $z$ and $y$ lattice directions are around the sites on which the monopoles reside
-- the center of the cubes of the original cubic lattice / the sites of the dual lattice, 
see Fig.~\ref{Fig:duallattice}.

We have now established the symmetry transformation properties for the
slowly varying fields in the solution of the gauge mean-field
Hamiltonian \eqref{H1}.  As we will describe in the next section this
allows us to make an explicit connection of these solutions to the
individual members in our family of dimer models.  In particular, we
directly show how the symmetries of the microscopic dimer interaction,
implicitly contained in the appropriate $F$ term in Eq.~\eqref{dual_QED},
re-enters the continuum field theory.  

\subsection{Effective Ginzburg-Landau actions and phase transitions}
\label{sec:action}

We will now turn to the individual members in our family of dimer models and derive an 
effective description in terms of a Ginzburg-Landau action that respects the symmetries
of the various models. 
This action is typically given in terms of the  two slowly varying complex fields $\phi_1$ and 
$\phi_2$ coupled to the dual U(1) gauge field $\alpha_{\mu}$. 
We then analyze the derived actions and discuss the nature of the phase transitions 
in these field theories.

Let us first establish some notations and introduce a three-component vector $\vec{N}$, 
which will serve as an order parameter indicating which dimer ordering is chosen as ground state
\be
\vec{N}(\phi_1,\phi_2) \equiv \phi^{\dagger}_{\alpha} \vec{\sigma}_{\alpha\beta} \phi^{\phantom\dagger}_{\beta}\;,
\label{eq:vectorN}
\ee
where $\vec{\sigma}$ are the three Pauli matrices. 
The six columnar ground states of our dimer models depicted in Fig.~\ref{Fig:DimerModels} then correspond to $\vec{N}$ pointing along positive or negative $x,y,z$ directions, respectively.
Finally, let $\vec{\phi} = \left( \phi_1, \phi_2 \right)$ be a two-component vector 
combining the two complex fields $\phi_1$ and $\phi_2$.

We can now write the effective Ginzburg-Landau action as
\be
{\mathcal S}_{\text{eff}} = \beta \int d^3r \left(\left|({\nabla}-i\vec{\alpha}) 
                     \vec{\phi}\right|^2 + U(\vec{N}) + {\mathcal L}(\vec{\alpha}) \right) \;,
\label{eq:action}
\ee
where the first $|\ldots|^2$-term is a minimal coupling of the two complex fields to the dual 
U(1) gauge field and ${\mathcal L}(\vec{\alpha})$ is the usual Maxwell's term for the gauge field.
The potential $U(\vec{N})$ is determined by the underlying symmetries of the various dimer
models, which are summarized in Table \ref{tab:symmetry}.
This potential therefore varies for the individual models as discussed in more detail
the following.

Note that the action \eqref{eq:action} does not contain any time-derivatives. 
The reason is that in the presence of such time-derivative terms and
periodic boundary conditions in imaginary time 
all modes  with non-zero Matsubara frequencies are more massive 
than the zero frequency mode of interest here and can be integrated out.

\begin{table}[t]
\begin{tabular}{c|c}
$\;\;$ dimer model $\;\;$ &  symmetries  \\
	\hline \hline
6-GS    & $T_x,\ T_y,\ T_z,\ R_{\frac{\pi}{2},Rxy},\ R_{\frac{\pi}{2},Rxz}$    \\ 	\hline
4-GS    & $T_x,\ T_y,\ T_z,\ R_{\frac{\pi}{2},Rxy}$                          \\ 	\hline
2-GS    & $T_x,\ T_y,\ T_z,\ R_{\frac{\pi}{2},Rxy}$                          \\ 	\hline
1-GS    & $T_x,\ T_y,\ R_{\frac{\pi}{2},Rxy}$                            \\ 	\hline \hline
xy         & $T_z,\ R_{\frac{\pi}{2},Rxy}$                                  \\	\hline
xxy       & $T_x,\ T_z$                                             \\	\hline
xyz       & $R_{\frac{\pi}{2},Rxy},\ R_{\frac{\pi}{2},Rxz}$  \\ \hline
xyzz      &  $T_z, \ R_{\frac{\pi}{2},Rxy}$                  \\ \hline
xxyyz     &  $T_x, \ T_y, \ R_{\frac{\pi}{2},Rxy}$     
\end{tabular}
\caption{Symmetries of the various dimer models.}
\label{tab:symmetry}
\end{table}

\subsubsection{The 1-GS model}
\label{sec:ana-1gs}

The ground state of the 1-GS dimer model is a single columnar ordering pattern,
which we choose to be oriented along the $z$ direction.
The effective Ginzburg-Landau action for this model is thus required to be invariant under the transformations $T_x$, $T_y$ and $R_{\frac{\pi}{2},Rxy}$ only.
In particular, the potential $U(\vec{N})$ in the action \eqref{eq:action} has the general form
\be
U(\vec{N}) = u_2 |\vec{N}| + v_2 N_z  = (u_2+v_2)|\phi_1|^2 + (u_2-v_2) |\phi_2|^2 
\;,
\label{eq:pot-1gs}
\ee
where we have only included terms up to quadratic order in the complex fields and
introduced two coupling constants $u_2$ and $v_2$. 
Note that the potential $U(\vec{N})$ introduces two inequivalent mass terms for the 
two complex fields.
As we reduce temperature it will be the complex field with smaller mass that
will condense thereby leading to a condensation of the monopoles 
(while the other field still has vanishing expecation value).
The corresponding phase transition can thus be described by a field theory with just one 
complex field coupled to a U(1) gauge field which is known to be a continuous 
transition in the inverted 3D XY universality class 
\cite{halperin:prl81}.
At this Higgs transition the system spontaneously breaks the U(1) gauge symmetry,
but does not break any lattice symmetries. 
As the Coulomb phase breaks down the charge excitations,  e.g. the monomers in the 
language of the dimer model, confine. 

We thus have clear analytical and numerical evidence for a continuous transition
between the Coulomb phase and a long-range ordered dimer crystal in this dimer
model, which cannot be explained by the standard LGW paradigm.

\subsubsection{The 2-GS and 4-GS models}
\label{sec:ana-42gs}

We now turn to the 2-GS and 4-GS models which we have seen to exhibit direct
first-order thermal transitions.
Although the two models have complementary ground-state manifolds, they are 
invariant under the exact same lattice transformations.
As given in Table \ref{tab:symmetry} these are the three lattice translations 
$T_x$, $T_y$, and $T_z$ as well as rotation around the $z$-axis, $R_{\frac{\pi}{2},Rxy}$.
The symmetry allowed  potential terms up to quartic order in the Ginzburg-Landau action 
\eqref{eq:action} are thus given by
\be
U(\vec{N}) = u_2 |\vec{N}| + u_4 |\vec{N}|^2 + v_4 N_z^2
\label{eq:act-4gs}
\;.
\ee

For the 4-GS model $v_4 > 0$, so the $N_z^2$ term modulates $\vec{\phi}$ by the constraint 
$|\phi_1|=|\phi_2|$ 
and prefers the order parameter $\vec{N}$ to point in the $xy$ plane. Noteworthily, this is still a continuously connected manifold with an internal U(1) symmetry for the order parameter. 
Note that the system does not need to break any lattice 
symmetries to satisfy this constraint (in contrast to the 6-GS model which we will discuss in
the next section). 
At the transition, when the complex fields $\vec{\phi}$ condense, the order parameter $\vec{N}$
becomes non-zero and points along one of the four lattice directions in the $xy$ plane.
Note that at this Higgs transition the system not only spontaneously breaks the U(1) gauge symmetry,
but {\em simultaneously} also the U(1) order parameter symmetry as well as the four-fold lattice symmetry.

The action \eqref{eq:act-4gs} exactly corresponds to the one studied for an easy-plane quantum 
antiferromagnet  in the context of deconfined quantum criticality \cite{senthil:sci2004,senthil:prb2004}.
While analytical investigations of this action have suggested a continuous phase transition
\cite{senthil:prb2004}, extensive numerical results have pointed to a weak first-order transition
\cite{Kragset:prl2006,jiang:2008}, which is also what we find in our present numerical analysis.

For the 2-GS model the order parameter $\vec{N}$ wants to point along the $z$ direction,
e.g. $N_z$ becomes maximal, which implies (contrary to the 4-GS model) that $v_4 < 0$.
This leaves the system with a disconnected manifold (of two points) for fixed magnitude 
$|\vec{N}|$ either prefering $|\phi_1|=0$ or $|\phi_2|=0$ and resulting in a $Z_2$ symmetry
for the order parameter.

In contrast to the 1-GS model, this theory cannot be reduced to a field theory with just one complex 
field {\em without} breaking the lattice symmetry $T_z$. One possibility now is to have {\em two}
subsequent transitions where we first break the lattice symmetry at a higher temperature and
subsequently observe a Higgs transition at a lower temperature (with an exotic intermediate phase 
of coexisting Coulomb and dimer crystal correlations). 
In terms of the complex fields $\vec{\phi}$ the system would spontaneously select one of the two
possibilities $|\phi_1| \neq 0, |\phi_2|=0$ or $|\phi_1| = 0, |\phi_2| \neq 0$ at the first transition.
At the second transition the non-vanishing $\phi$-field would require a fixed phase in a Higgs 
transition.

Another possibility is to have one direct transition. However, it is hard to imagine a field theory 
giving rise to a continuous transition
where the spontaneous breaking of the discrete $Z_2$ order parameter symmetry occurs 
simultaneously with the Higgs transition breaking the U(1) gauge theory. 
Thus, we conclude that a {\em direct} transition
is likely first-order. It appears to be the latter scenario that we observe in our numerics.

Since we have only numerical (and not analytical) evidence for a first-order transition in the 
4-GS model, but have some analytical evidence for a first-order transition in the
2-GS model, we probably expect the latter to be the stronger first-order
transition. This is what we observe in the numerical simulations of Sec.~\ref{sec:numerics}.

\subsubsection{The 6-GS model}
\label{sec:ana-6gs}

Finally, we turn to the 6-GS model which respects all the cubic lattice symmetries.
In writing down a symmetry allowed potential for the action \eqref{eq:action} we
consider the simplest invariants of the form
\be
 U(\vec{N}) = V(|\vec{N}|) + v_8 I_8(\vec{N}) \,,
 \label{eq:6GS-action}
\ee
where $I_8(\vec{N}) = N_x^2N_y^2 + N_y^2N_z^2 + N_x^2N_z^2$ is an eighth order
term in the complex fields $\vec{\phi}$ and we adopted a notation similar to
the one of Ref.~\onlinecite{senthil:prb2005}.
Again we can expand the potential term $V(|\vec{N}|) = u_2 |\vec{N}| + u_4 |\vec{N}|^2 + \ldots $.
Omitting the 8th order term in \eqref{eq:6GS-action} gives an SU(2) invariant action,
which has attracted some interest due to recent proposals of SU(2) invariant deconfined quantum 
critical points, as suggested in the $J-Q$ quantum model \cite{sandvik:prl07,melko:prl08,jiang:2008}.

Following the line of arguments in Ref.~\onlinecite{senthil:prb2005} the confining Higgs
transition (which we observe) occurs for $u_2 < 0$ and $|\vec{N}|$ {\em simultaneously}
acquires a finite magnitude, e.g. the U(1) gauge symmetry and the lattice symmetry are 
broken at the same transition.
As argued in Ref.~\onlinecite{senthil:prb2005} one of the six columnar ground states is selected by the eighth order term $v_8 I_8(\vec{N})$ with $v_8>0$. Since the Higgs transition of the action \eqref{eq:action} is suggested to be continuous without this 8th order term and this term is likely irrelevant (due to its high order), we conclude that the action
\eqref{eq:6GS-action} allows for a continuous transition. 
This seems to be in agreement with the numerical evidence of Ref.~\onlinecite{alet:prl2006} and our 
present numerical analysis interpolating between the 4-GS and 6-GS models.

Direct numerical simulation of the SU(2) invariant action \eqref{eq:6GS-action} without the 8th order 
term has provided controversial results with some evidence for a continuous transition 
\cite{motrunich:08}, while another recent analysis favors a weak first-order transition \cite{kuklov:prl08}. 
Including the 8th order term in the action \eqref{eq:6GS-action} a direct numerical simulation of 
a similar action to \eqref{eq:6GS-action} reported a continuous transition with exponents close but apparently different from those of the 3D $XY$ model \cite{charrier:prl08}.

\subsubsection{The interpolated models}

Finally, we briefly turn to the models interpolating between models exhibiting continuous and first-order 
transitions as discussed in section \ref{sec:interpolation}.

Interpolating between the 1-GS and 2-GS model the system exhibits for all interpolation parameters
$0 \le \lambda < 1$ the same lattice symmetries and is therefore described by the same
Ginzburg-Landau action with potential \eqref{eq:pot-1gs} as the 1-GS model. 
This symmetry analysis suggests that for all $\lambda<1$ the two complex fields $\phi_1$, $\phi_2$
acquire different masses and we can describe the action in terms of a {\em single} complex field 
coupled to a U(1) gauge field. On the other hand, we expect the strong first-order transition of the
2-GS model ($\lambda=1$) to be stable towards small perturbations and therefore to extend over
a finite region $\lambda<1$. As a consequence, there should be a multicritical point $\lambda_c$ 
where the line of continuous transitions for $\lambda \le \lambda_c$ meets the first-order line for $\lambda>\lambda_c$. 

Interpolating between the 4-GS and 6-GS model the system exhibits for all interpolation parameters
$0 \le \lambda < 1$ the same lattice symmetries, while the symmetries change for the endpoint 
$\lambda=1$ which corresponds to the 6-GS model, see also Table~\ref{tab:symmetry}. 
Again this symmetry analysis suggests that all interpolated models with $0 \le \lambda < 1$ are described by the same Ginzburg-Landau action with potential \eqref{eq:act-4gs}. This seems to
be in agreement with our numerical results suggesting that all models with $0 \le \lambda < 1$
exhibit first-order transitions.


\section{An Intermediate Paramagnet}
\label{sec:variation}

Finally, we turn to a second family of dimer models that also energetically favor specific subsets of
the six columnar ordering patterns illustrated in Fig.~\ref{Fig:DimerModels}.
The distinct feature of this second family of dimer models is that they harbor {\em two consecutive}
thermal phase transitions. The high-temperature phase transition out of the Coulomb phase is
again driven by the condensation of monopoles with confining monomer excitations. However, 
this phase transition is into a paramagnetic phase without dimer crystalline order which only 
forms at the low-temperature transition. Thus, we are left with an unusual sequence of phases
in these models with the paramagnet residing at {\em intermediate} temperature scales.

\subsection{A second family of dimer models}

\begin{figure}[t]
\includegraphics[width=\columnwidth]{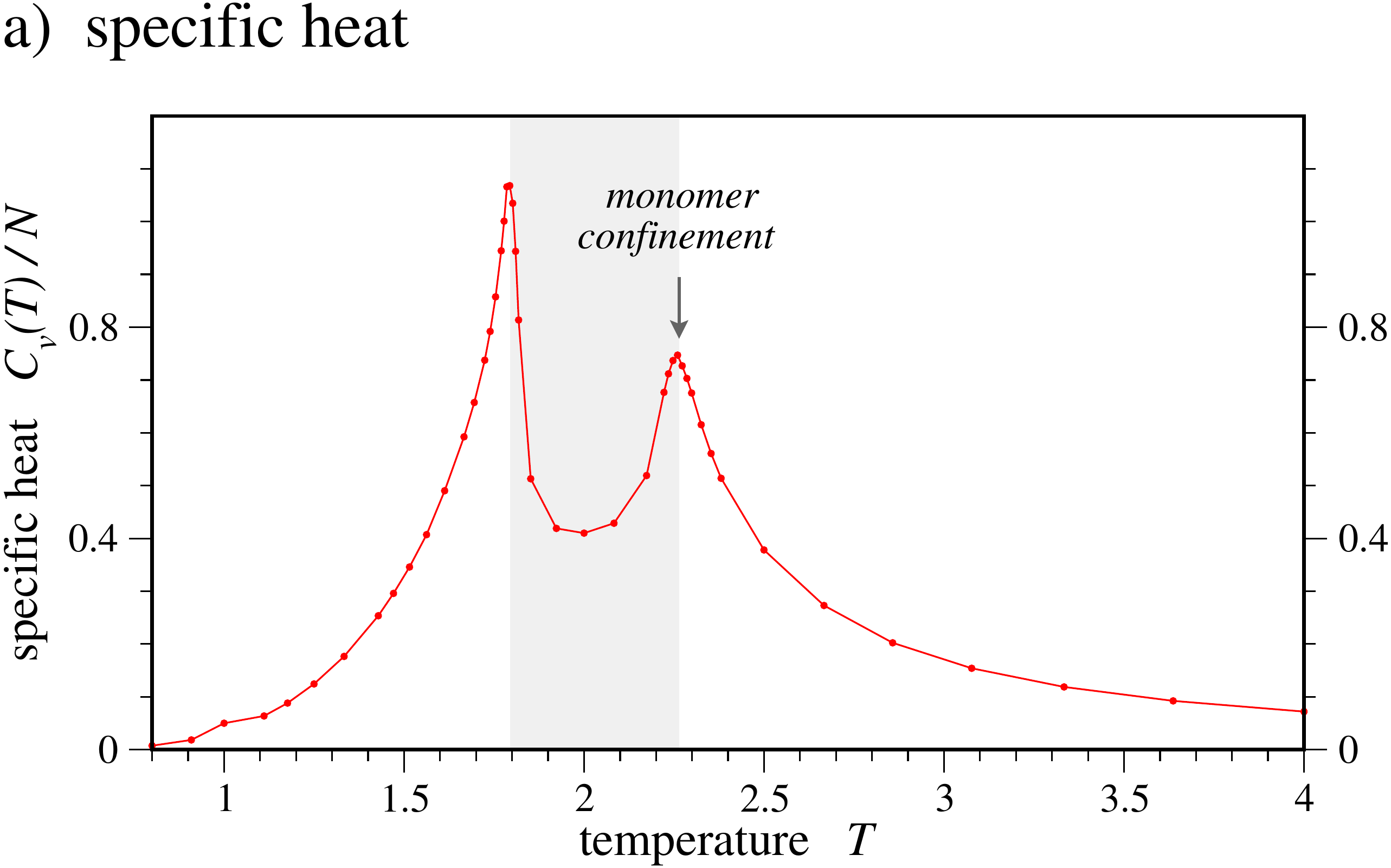}
\vskip 4mm
\includegraphics[width=\columnwidth]{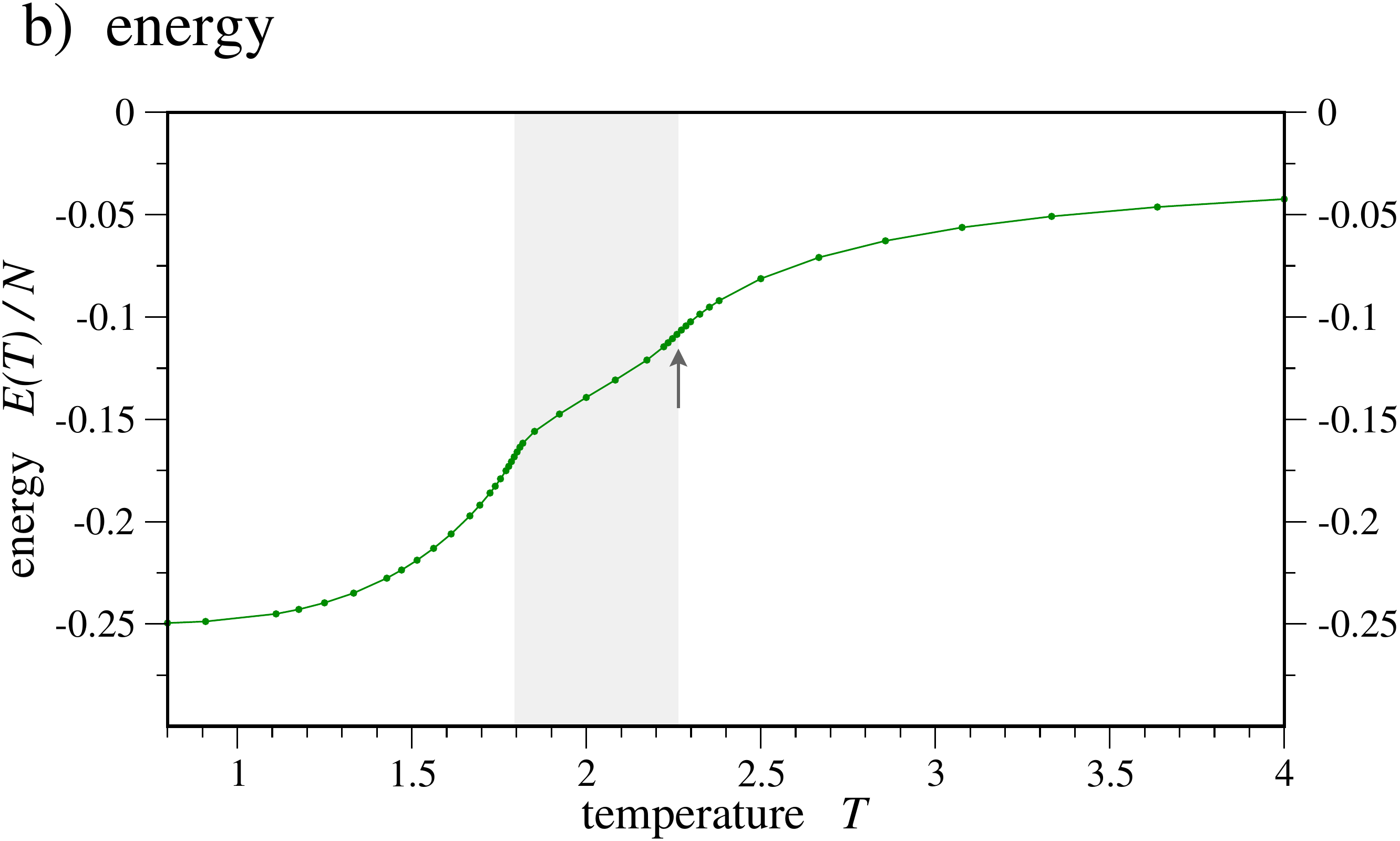}
\vskip 4mm
\includegraphics[width=\columnwidth]{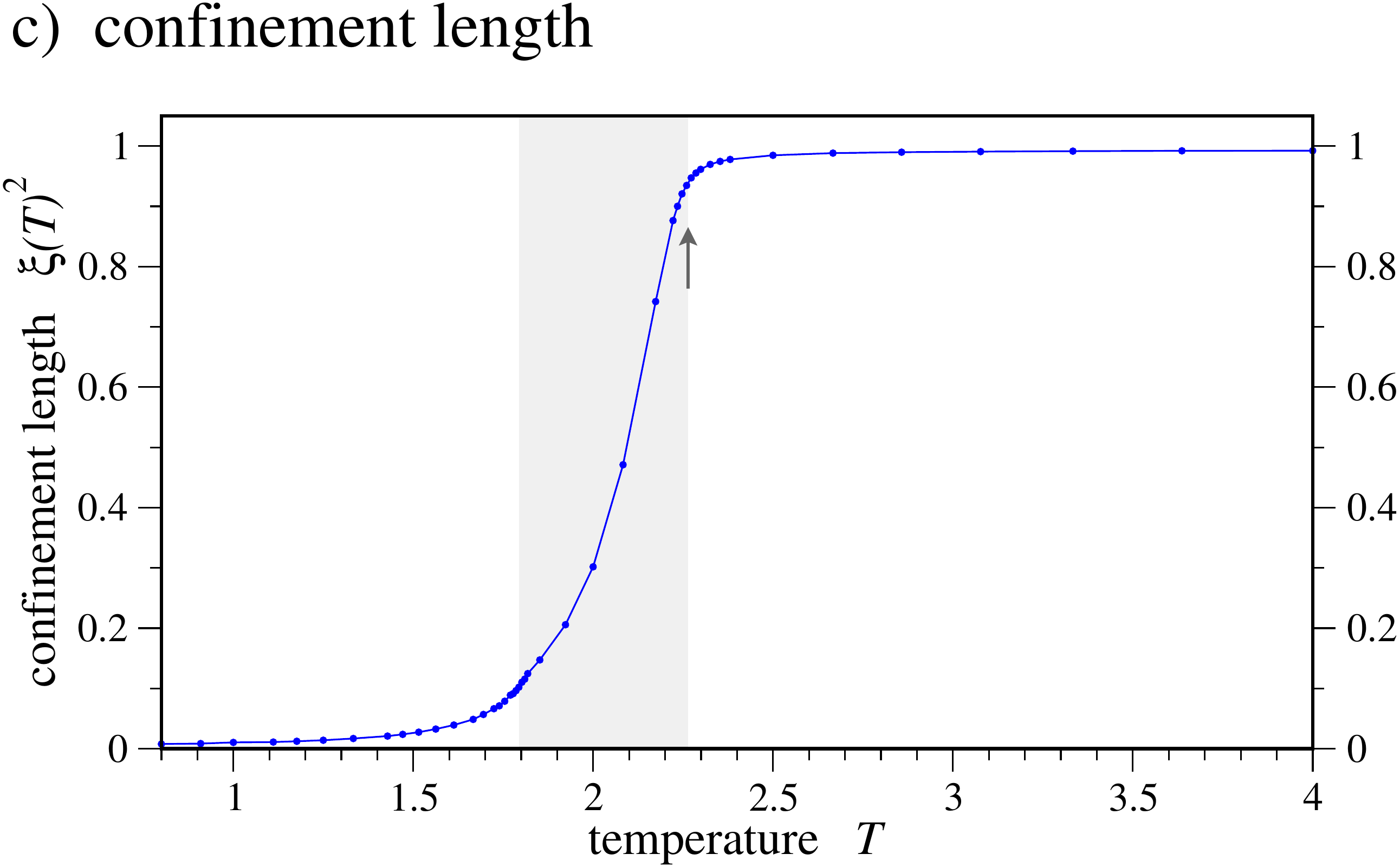}
\caption{
  (color online) Overview of the xy-model that favors one columnar ordering pattern along
  the $x$ and $y$ lattice directions: 
  a) specific heat per site $C_v(T)/N$, b) energy per site $E(T)/N$, and c) confinement length $\xi^2$.
  This model undergoes an unusual sequence of phases with a paramagnetic phase
  (shaded area) at intermediate temperature scales.
  At the high-temperature transition out of the Coulomb phase (denoted by an
  arrow) the monomers confine,
  while the dimer crystal forms only at the low-temperature phase transition.
}
\label{Fig:xy-overview}
\end{figure}

Our second family of dimer models explores other combinations of the six columnar ordering
patterns in Fig.~\ref{Fig:DimerModels} as ground states. 
The common characteristic in selecting the admissible ground states is that for at least one lattice 
direction we choose only one of the two possible columnar orderings and there is more than one 
ground state. If we name the models by the lattice directions for which ground states are chosen,
these are the `xy', `xyz', `xxy', `xyzz' and `xxyyz' models.
In this nomenclature the models in our first family of models would be named 'z','zz','xxyy' and 
'xxyyzz' for the 1-GS, 2-GS, 4-GS and 6-GS model, respectively.

We will not discuss all possible models in this section, but concentrate on the `xy' model with Hamiltonian
\be
{\mathcal H}_{xy}   =  - \sum_{\substack{\square}} (n^e_{=} + n^e_{//})   \;,
\label{eq:xy-model}
\ee
where we have chosen the columnar dimer orderings on the even bonds in the $x$ and $y$
lattice directions as ground states.

\begin{figure}[t]
\includegraphics[width=\columnwidth]{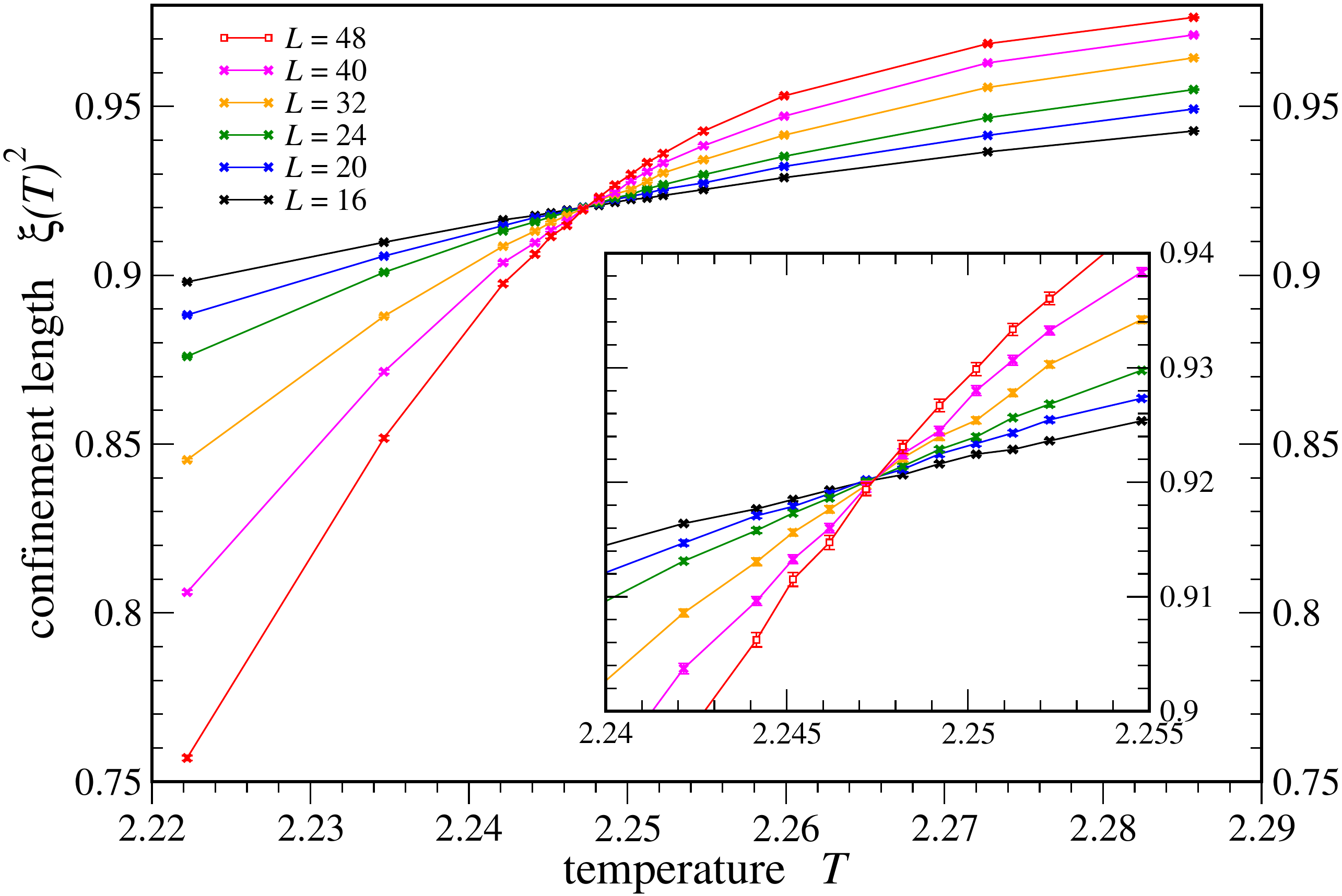}
\caption{
  (color online) 
  Crossing point of the confinement length $\xi^2(T)$ for different system sizes at the high-temperature 
  phase transition out of the Coulomb phase for the xy model.
}
\label{Fig:xy-Confinement}
\end{figure}

We summarize our numerical results for this model in Fig.~\ref{Fig:xy-overview}. 
The two consecutive thermal transitions both carry distinct thermodynamic signatures with
a double peak structure emerging in the specific heat. 
At the high-temperature transition out of the Coulomb phase the monomer confinement length
drops again indicating that this transition is due to monopole condensation. A finite-size scaling
analysis reveals a distinct crossing point (see Fig.~\ref{Fig:xy-Confinement}) indicating a 
continuous transition into the intermediate temperature paramagnet. We will argue that this
transition is again described by the inverted 3D XY universality class. 

\begin{figure}[t]
\includegraphics[width=\columnwidth]{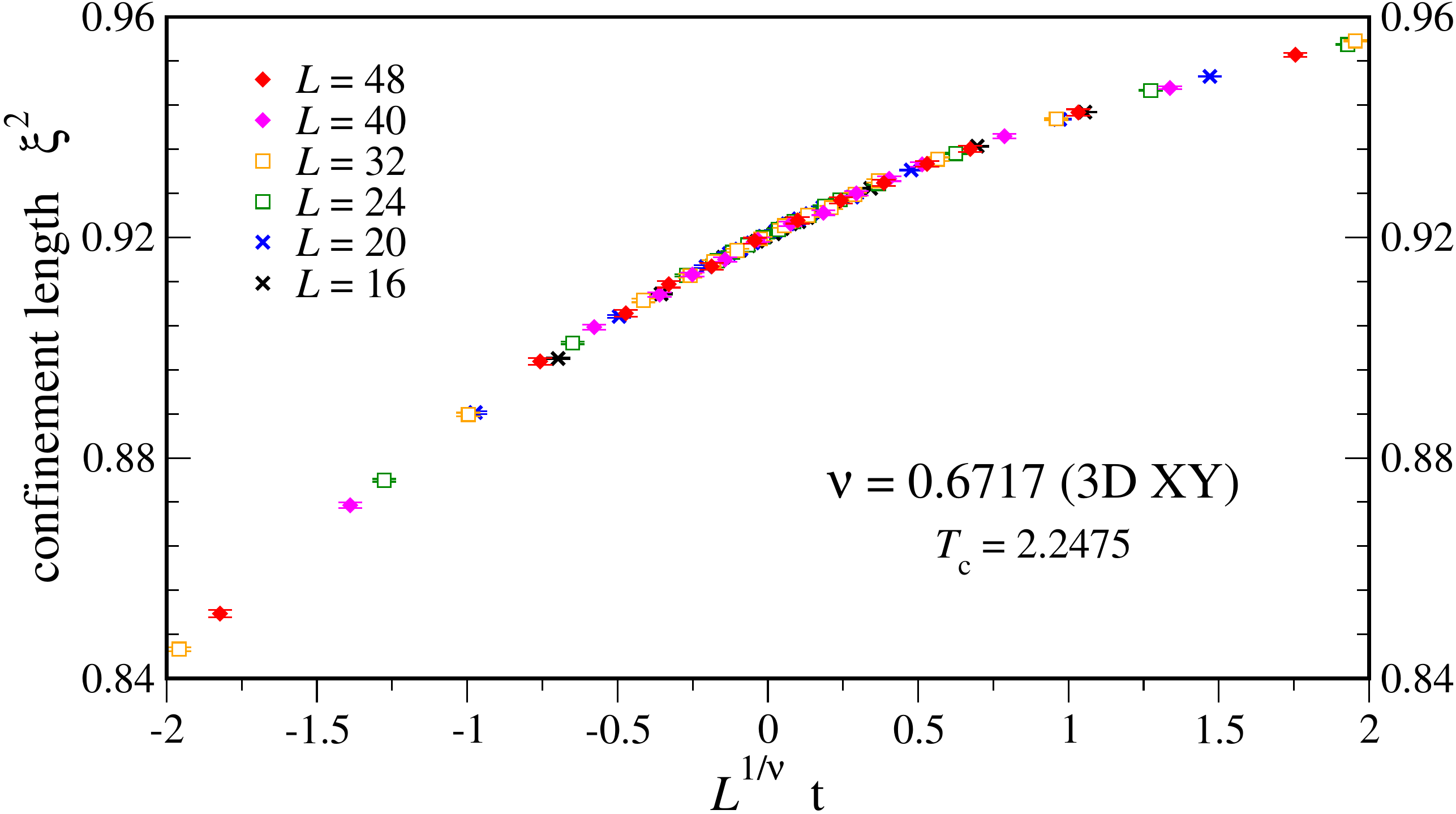}
\caption{
  (color online) 
  Data collapse for the confinement length $\xi^2$
  as a function of $L^{1/\nu} t$,  where $t = (T-T_c)/T_c$ with $T_c = 2.2475$.
  The critical exponent $\nu = 0.6717$ corresponds to the 3D XY universality class.
}
\label{Fig:xy-DataCollapse}
\end{figure}

We first notice that that the temperature of the Coulomb transition in the xy model ($T_c\simeq 2.247$) 
turns out to be close to the one found for the 1-GS model (where $T_c \simeq 2.276$). 
Another indicator that the Coulomb transitions in these two models are closely related and 
probably of the same universality class is that the universal value of the confinement 
length at the crossing point is $\tilde{\xi}(T_c) \approx 0.920 \pm 0.001$, which is rather close 
to the one found for the 1-GS model ($\tilde{\xi}(T_c) \approx 0.923 \pm 0.001$).
As a final argument, we find an excellent data collapse of the confining length measured in the vicinity of this transition for different system sizes when rescaling the data with the correlation length exponent $\nu=0.6717$ of the 3D XY universality class, see Fig.~\ref{Fig:xy-DataCollapse}.

\begin{figure}[t]
\includegraphics[width=\columnwidth]{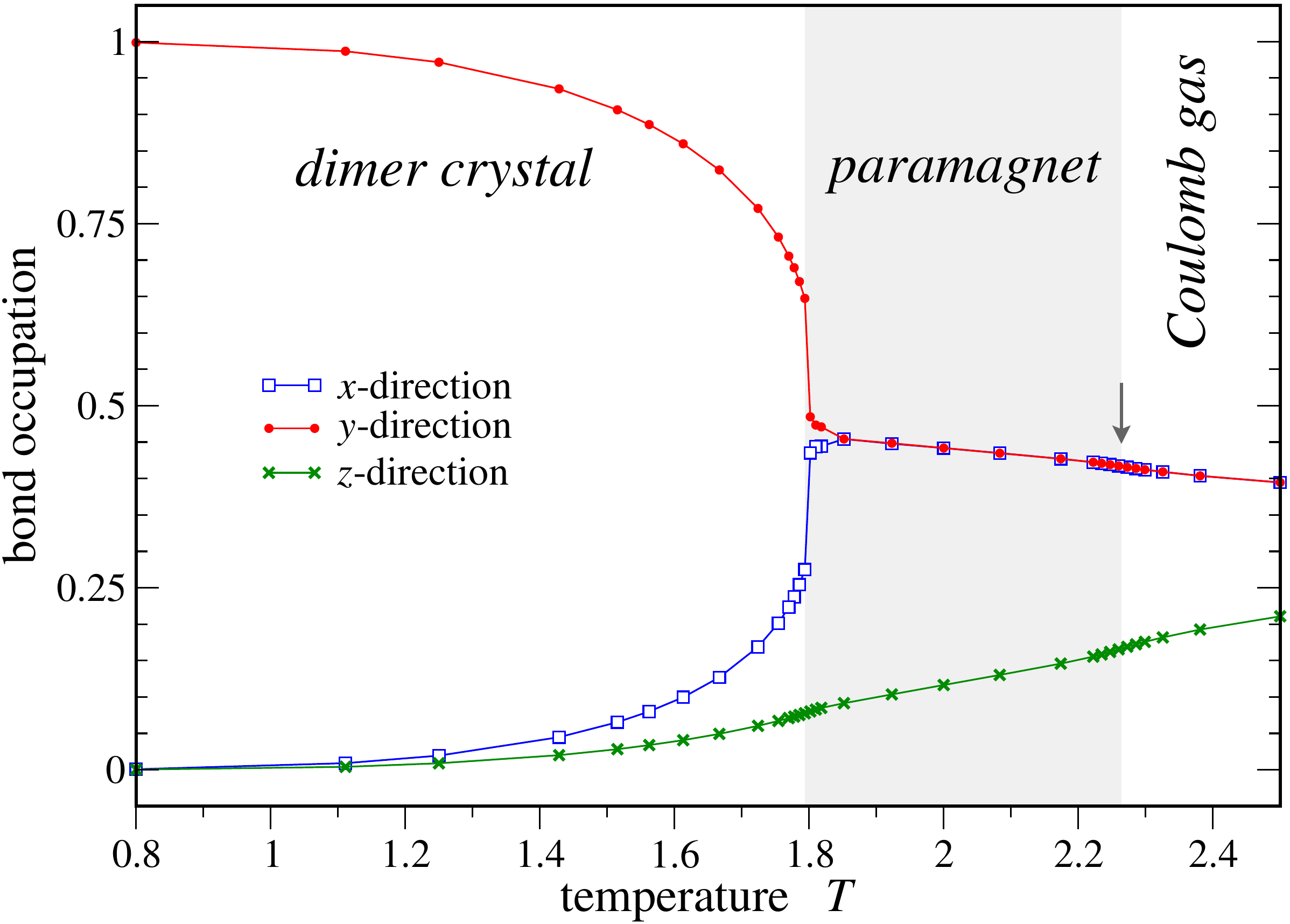}
\caption{
  (color online) 
  Bond occupation along the $x$, $y$ and $z$ lattice directions versus temperature for the xy model.
  At low temperatures data is shown for a system that spontaneously orders along the $y$ direction 
  (other systems spontaneously order along the $x$ direction). 
  The low temperature transition between the paramagnet and the dimer crystal is accompanied by
  a sudden increase of the bond occupation along the preferred lattice direction. 
  The high-temperature transition between the Coulomb gas and the paramagnet where the
  monomers confine is indicated by the arrow.
}
\label{Fig:xy-OrderParameter}
\end{figure}

\begin{figure}[t]
\includegraphics[width=\columnwidth]{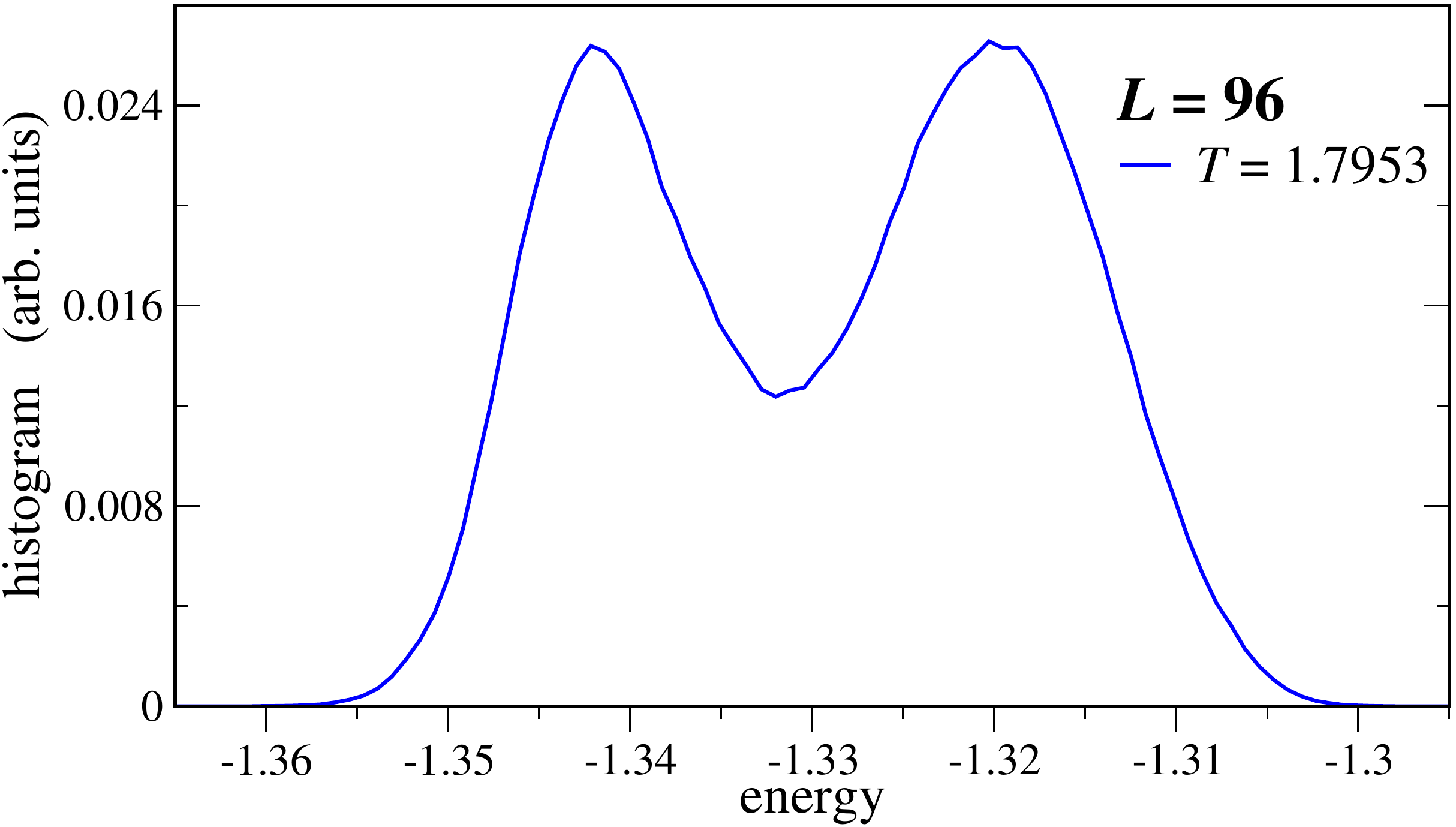}
\caption{
  (color online) 
  Energy histogram in the vicinity of the low-temperature phase transition,
  which we argue to be a thermal analog of a spin-flop transition (see text).
  Our numerics indicate a weak first-order transition by a double-peak structure
  in the energy histogram, which emerges for larger system sizes. 
  Here system size is $L=96$.
}
\label{Fig:xy-Histogram}
\end{figure}

At the low-temperature transition the system spontaneously selects one of the 
two possible columnar ordering patterns and we observe a sudden increase 
of the number of plaquettes with parallel dimers, e.g. $n^e_{=}$ or $n^e_{//}$, 
respectively (see Fig.~\ref{Fig:xy-OrderParameter}).
The sharp jump of this order parameter indicates a
likely occurence of a first-order phase transition. The first-order nature
of the low-temperature phase transition is indeed confirmed by the bimodal
structure of energy histograms close to this transition point (see
Fig.~\ref{Fig:xy-Histogram}). 


\subsection{Theoretical Analysis}
\label{sec:the-var}

We can discuss the nature of the phase transitions for this second type of dimer models 
by again analyzing the symmetry-allowed effective Ginzburg-Landau actions in full analogy 
to the discussion in section \ref{sec:action} for the first family of dimer models.

With the lattice symmetries for the individual members of this second family of dimer models
given in Table~\ref{tab:symmetry}, we find the following potentials for the Ginzburg-Landau 
action in Eq.~\eqref{eq:action}

\bea
U(\vec{N}) =
\left\{
\begin{array}{ll}
u_2 |\vec{N}| + v_2 (N_x+N_y),  & \textrm{xy model},  \\
u_2 |\vec{N}| + v_2 N_y,  & \textrm{xxy model},      \\
u_2 |\vec{N}| + v_2 (N_x+N_y+N_z),  & \textrm{xyz model}, \\
u_2 |\vec{N}| + v_2 (N_x+N_y),  & \textrm{xyzz model}, \\
u_2 |\vec{N}| + v_2 N_z ,  & \textrm{xxyyz model}
\;.
\end{array}
\right.
\label{eq:pot-var}
\eea

For all models we can diagonalize the quadratic part in these potentials
by performing a SU(2) rotation in the $(\phi_1,\phi_2)$ space such that
the potentials \eqref{eq:pot-var} take an identical form as given in
Eq.~\eqref{eq:pot-1gs} for the 1-GS model.  As a consequence, we expect
the high-temperature transition out of the Coulomb phase in all these
models to be described by the same Higgs mechanism we identified for the
1-GS model resulting in a continuous transition in the inverted 3D $XY$
universality class.
Our numerics give supporting evidence for a continuous transition in this universality class
as discussed above.

The main distinction between the models in our second family of models
and those in the first set of models is that here we can break an
additional lattice symmetry which apparently gives rise to the second
transition into the dimer crystal phase at lower temperatures.  We argue
that this second low-temperature phase transition is generically a
first-order transition analog to a {\em spin-flop} transition.  To see
this analogy consider the xy model with the potential $U(\vec{N}) = u_2
|\vec{N}| + v_2 (N_x+N_y)$ in the Ginzburg-Landau action.  Below the
confinement transition the order parameter $\vec{N}$ has a non-zero
expectation value (since the monopoles are condensed), but points
half-way between the $x$ and $y$ directions, thereby minimizing the
second term in the potential. This is also evident in our numerical
simulations as shown in Fig.~\ref{Fig:xy-OrderParameter}.  At very low
temperatures, however, we know that the system must (because there will
be no dimer fluctuations) spontaneously order along one of the two
lattice directions, thus breaking the symmetry between $x$ and $y$
directions.  Therefore the spin $\vec{N}$ must reorient away from the
$(110)$ axis to the $(100)$ or $(010)$ axis.  To describe this, we
require additional higher-order terms in the potential $U(\vec{N})$, of
the form $N_x N_y$, $N_x^2 N_y^2$, etc.  At low temperatures, since the
magnitude of the spin becomes large ($|N|=1$ as there are no
fluctuations), such terms are no longer negligible.  On lowering the
temperature and increasing these higher order terms, we expect that the
minimum directions of this energy function may abruptly switch to their
low temperature values.  This is indeed the most commonly occurring
situation in spin systems, in which such a first-order reorientation is
known as a ``spin flop'' transition.  This expectation, arrived at above
from ``analytical'' field theory considerations, is indeed verified in
the numerics (see Fig.~\ref{Fig:xy-Histogram}).


\section{Discussion}
\label{sec:discuss}

Recent years have seen an extensive search for continuous phase transitions
beyond the LGW paradigm, which were originally suggested to occur in certain
quantum models \cite{senthil:prb2004,senthil:sci2004}.
In this manuscript, we have demonstrated that this exotic physics can 
manifest itself also in various {\sl classical} models. This, of course, is not much of a surprise
since the universality of continuous phase transitions mandates that they occur
in a large variety of models, including classical ones. 
Nevertheless, it is amusing to note that such unconventional phase transitions and 
the sophisticated ordering mechanisms associated with them can actually be found 
in simple variations of one of the golden models of statistical mechanics, 
namely the dimer model. 
The key ingredient giving rise to this exotic physics is a constraint which enforces 
close-packed coverings of hard-core dimers. In a way, this readily builds 
into the classical model a certain level of frustration which is often invoked to be a key 
ingredient for quantum models to exhibit non-LGW criticality.

Besides the important step to directly establish the occurrence of non-LGW transitions
in these dimer models, we view several advantages arising from their classical nature:
(i) Classical models are notoriously simpler to analyze, both theoretically and
numerically, than quantum models. They are accessible to Monte Carlo
approaches, thus allowing to study critical phenomena through the direct
simulation of large systems.
For the specific dimer models at hand, the existence of a highly-efficient Monte 
Carlo worm algorithm is also very attractive.
(ii) These models ease the identification of the necessary ingredients that are 
needed to cause non-LGW physics in a lattice model (such as lattice and/or
continuous symmetries). 
This further opens the possibility of `reverse-engineering'
or `rolling back the path integral' 
 to obtain two-dimensional quantum models that exhibit the same
non-LGW criticality as their three-dimensional classical counterparts. 
Such a classical-to-quantum mapping was recently used in Ref.~\onlinecite{powell:prl08}.
(iii) The stability of certain critical behavior can be easily explored in variations
of these classical models, e.g. through the inclusion of perturbations or by
extrapolating terms (as performed in the current study).
For instance, a yet-to-be-explored possibility is to include terms that {\it frustrate} 
the columnar ordering. A similar situation in a quantum model would generically 
come with a sign problem in Quantum Monte Carlo simulations, putting serious 
limitations to any numerical study.
(iv) Finally, the sheer simplicity of these models might indicate that non-LGW transitions
are not that exotic after all.


\section{Acknowledgments}\
Our numerical work used some of the ALPS libraries \cite{ALPS,Troyer},
see also http://alps.comp-phys.org.  
This work was supported by the DOE through Basic Energy Sciences grant
DE-FG02-08ER46524. LB's research facilities at the KITP
were supported by the National Science Foundation grant NSF PHY-0551164.


\end{document}